\documentclass[10pt,twocolumn,letterpaper]{article}

\usepackage[margin=1in]{geometry}
\usepackage{authblk}
\usepackage{hyperref}
\usepackage{float}
\usepackage{graphicx}
\usepackage{amsmath}
\usepackage{dblfloatfix} 
\usepackage{flushend}    
\usepackage{amssymb}

\begin{document}

\title{Embodying Intelligence into Mechanical Metamaterials via Reservoir Computing}

\author[a]{Shan He}
\author[b]{Steven Kiyabu}
\author[c]{Philip R. Buskohl}
\author[a,1]{Patrick Musgrave}

\affil[a]{Mechanical and Aerospace Engineering Department, University of Florida, Gainesville, FL 32611}
\affil[b]{UES, a BlueHalo company, Dayton, OH 45432}
\affil[c]{Materials and Manufacturing Directorate, Air Force Research Laboratory, Wright-Patterson Air Force Base, OH 45433}


\maketitle

\begin{abstract}
This study harnesses the embodied intelligence of mechanical metamaterials to sense and process environmental vibrations with minimal digital computation. Using physical reservoir computing (PRC), we turn the metamaterial and its nonlinear dynamics into a physical neural network that nonlinearly transforms the input vibrations and uses a simple linear training to compute a range of tasks. We introduce a novel metamaterial reservoir composed of a network of unit cells with contact nonlinearities that are the physical equivalent of leaky rectified linear unit (ReLU) activation functions. We experimentally show that the metamaterial reservoir can compute two classes of tasks: independent tasks—such as benchmark functions—and embodied tasks—such as proprioception—which we introduce to describe tasks coupled to the structure's dynamics. By comparing against a linear metamaterial, we demonstrate that nonlinearity is critical for high task performance, and we show that the metamaterial is robust to inputs of varying complexity. Through a dimensionality reduction, we uncover the governing information separation mechanism and show that the metamaterial separates the input vibrations into new frequency content spatially distributed across the sensor readouts. We then confirm that frequency content is a key indicator of task performance by conducting an optimal sensor selection study using a frequency-based greedy algorithm. Finally, we demonstrate that a metamaterial’s generalized performance for different tasks can be quantified using the memory vs. nonlinearity subspace, providing a design tool for other reservoir abstractions. These results establish the embodied intelligence of mechanical metamaterials and provide a path for sense-assess-response in intelligent systems.
\end{abstract}

\vspace{1em}
\vspace{1em}

\section*{Significance Statement}
This study demonstrates that mechanical metamaterials can function as physical computers, processing vibrations through their structural dynamics rather than electronics. We show for the first time that a metamaterial can perform diverse computational tasks from benchmark functions to embodied tasks such as proprioception, where we define embodied tasks as tasks coupled to the structure’s own dynamics. Using a novel metamaterial with nonlinearities resembling mechanical activation functions, we reveal the fundamental computational mechanism as the mechanical separation of input vibrations into frequency components distributed across the structure. We establish design principles by relating metamaterial nonlinearity and memory to computational performance. This work provides a foundation for embodying sensing and computation into physical structures, enabling intelligent structures that respond to their environment.

\section*{Introduction}
Biological systems demonstrate embodied intelligence where they can sense, assess, and respond to external stimulation from the environment intrinsically through the natural dynamics of their physical bodies. For example, birds often change the shapes of their wings in order to perform different flight maneuvers and tasks \cite{tobalske2007biomechanics}. This enables decision-making to occur in real time, with efficient energy usage and robust adaptability. Inspired by biological systems, engineered systems such as soft robots \cite{wang2024advancements, eyvazian2026state, rus2015design}, morphing wings \cite{mowla2025recent, ameduri2023morphing}, and smart materials \cite{hoffmann2023bird, ma2019review} have sought to perform specific tasks, such as adaptive locomotion, shape morphing, and sensing environments, efficiently and effectively. These effects reflect a pursuit of embodied intelligence where sensing, computation, and response are integrated within the physical bodies of the systems \cite{kortman2025perspectives}. However, these bio-inspired engineered systems usually require large information processing through centralized computers \cite{aner2025decade, kim2013soft} due to highly nonlinear dynamics inherent to compliant or deformable structures, leading to energy-usage overhead and delay in control.

Recent advances in Physical Reservoir Computing (PRC) have shown that harnessing the nonlinear dynamics of a physical system as a computational resource provides a pathway to achieve embodied intelligence \cite{nakajima2020physical,tanaka2019recent}. In Reservoir Computing (RC), a fixed high-dimensional nonlinear dynamic system, called the reservoir, transforms input information into rich separated signals, requiring only a linear readout layer that is trained to perform a desired task \cite{jaeger2004harnessing,maass2002real}. PRC extends this concept by embodying the intelligence directly into a physical material or structure \cite{nakajima2020physical}, analogous to biological systems. In PRC, information is processed through the system's natural dynamics rather than through digital computation on conventional in-silico hardware. By doing so, PRC offloads the majority of the information processing onto the physical system itself, enabling computation with minimal digital energy, mitigating control-latency, and reducing the need for modeling of complex dynamics \cite{tanaka2019recent}. PRC can be implemented across different types of physical systems or devices including electronic circuits \cite{liang2024physical, nowshin2020recent}, optical setups \cite{rafayelyan2020large, bu2022efficient}, fluid reservoirs \cite{vincent2025fluid, vincent2026information, goto2021twin}, and mechanical structures \cite{kiyabu2025optomechanical, zhang2022harnessing, bhovad2021physical}. Among these platforms, recent studies have shown that mechanical deformation itself can serve as a computational resource for PRC, with information embedded through the system's nonlinear dynamic response \cite{nakajima2013soft, coulombe2017computing, he2025physical}.

Mechanical metamaterials are architected structures designed to achieve unique mechanical behaviors that are not typically attained in natural or homogeneous materials. Recent research has shown that mechanical metamaterials have information processing capabilities and can act as mechanical computers \cite{yasuda2021mechanical}. Mechanical structures can perform mechanical logic \cite{raney2016stable} and programmable shape deformation \cite{coulais2016combinatorial}. Metamaterials can also be designed to perform specific time-varying tasks such as adaptive band-gap tuning \cite{wang2025metamaterial, zhang2023embodying}, wave cloaking \cite{bordiga2025nonlinear}, and shape morphing \cite{poon2019phase}. Further, by embedding microprocessors into the structure, a metamaterial can become a mechanical neural network that can execute programmed tasks \cite{lee2022mechanical, haghpanah2016programmable}. However, there is a need for mechanical metamaterials that can perform a broad range of time-varying computational tasks without embedded processors.

Physical Reservoir Computing shows promise for realizing mechanical metamaterials with information processing capabilities; however, there is a need to fundamentally understand the physical mechanisms through which a metamaterial reservoir computer separates information. Further, there is a need to understand the connection between this information separation process and the role of the readout layer, including how informative is distributed spatially across the readouts.

Previous work has explored the role of dynamic modes in task performance \cite{he2025role}, but was limited to a 1-D structure that requires large dimensionality for performance and did not consider readout selection or task generality. Another study \cite{he2025physical} has explored sensor selection but only as a basic ranking and offered no mechanistic understanding of why certain readouts were preferred.

In this study, we establish a metamaterial reservoir computer capable of processing information through its intrinsic nonlinear dynamics to demonstrate the role of nonlinearity in computing two different classes of state-estimation tasks. This study introduces a vibration-based metamaterial structured as a network of nonlinear beam unit cells, each exhibiting a leaky rectified linear unit (ReLU)-like bending stiffness that serves as a physical activation function. Using the PRC framework, we demonstrate that the metamaterial has high computational capabilities to perform a set of application-relevant tasks, including an embodied task that predicts the strain rate at a surface location and an embodiment-independent task involving emulation of a ReLU-based benchmark function. We show that the metamaterial has robust prediction performance across a range of input complexities. We further quantitatively reveal that nonlinear frequency separation is the governing physical mechanism, enabling information processing in the mechanical reservoir. We identify the optimal set of sensors that spans the network's dynamic space and captures the most relevant information for computation of a given task. Finally, we analyze how computation emerges from the interplay between nonlinearity and memory by mapping each task and individual sensors within a unified nonlinearity–memory metric space.

The paper is organized as follows. We first present an overview of the nonlinear metamaterial reservoir and demonstrate its performance on both embodied and independent tasks. We then demonstrate performance with tasks of increasing input complexity, and we validate the metamaterial's effectiveness by comparing against the linear counterpart. Next, we reveal the underlying nonlinear encoding mechanism by examining the frequency separation using frequency content analysis in conjunction with Principal Component Analysis for visualization purposes. We then evaluate the readout–task compatibility and guide sensor reduction using a frequency alignment method. Finally, we quantify nonlinearity and memory of the reservoir’s dynamics and a series of tasks within a unified metric space.

\section*{Metamaterial Reservoir Computer}
\begin{figure*}[tb!]
    \centering
    \includegraphics[width=\textwidth]{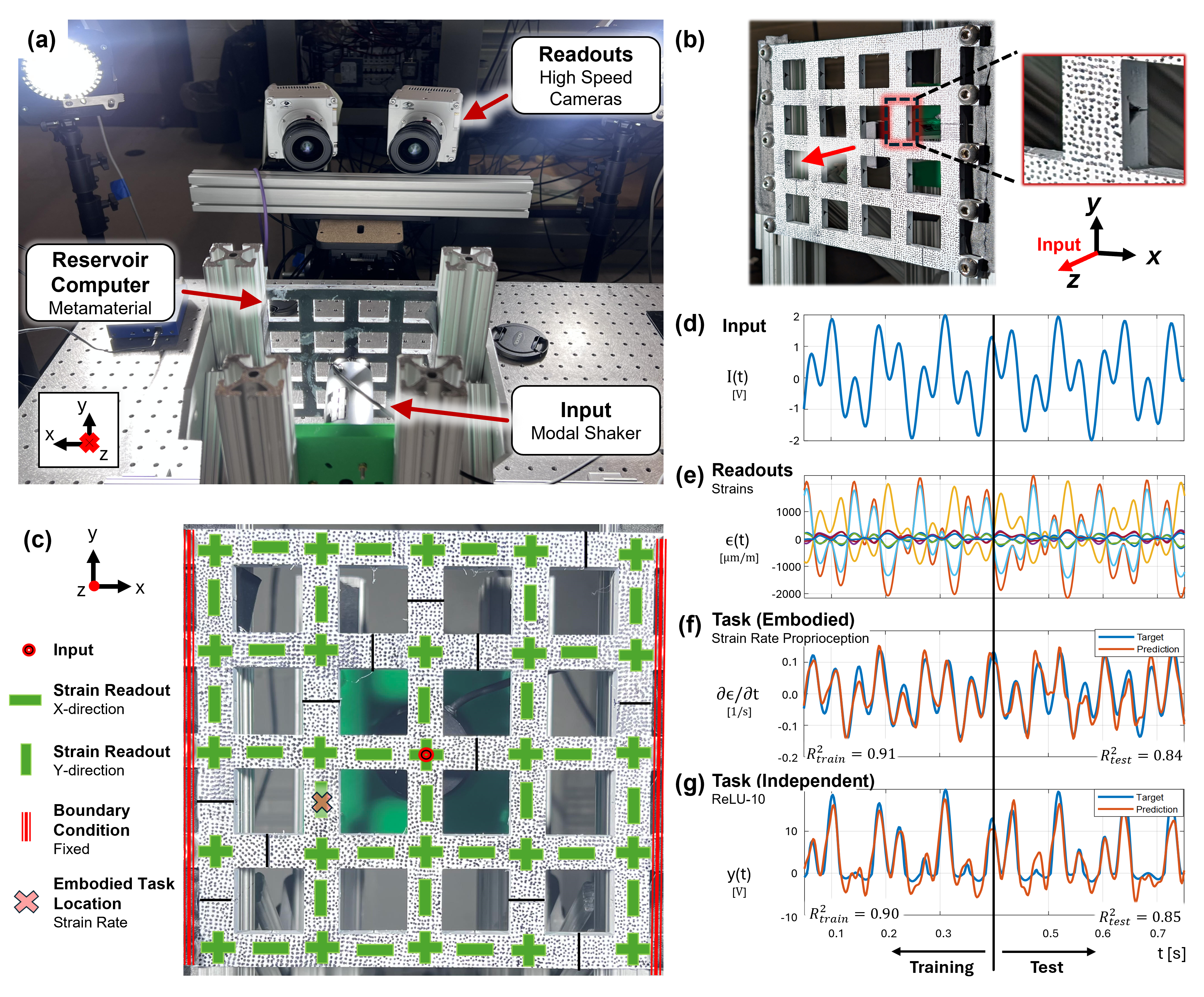}
    \caption{Overview of a nonlinear vibratory mechanical metamaterial functioning as a physical reservoir computer. (a) Experimental setup: A linear modal shaker is used to exert a force input. The nonlinear metamaterial is fixed in a clamped-free-clamped-free boundary condition (clamped on horizontal edges and free on vertical edges) and serves as the physical reservoir. Stereo DIC captures the dynamic response of the metamaterial reservoir readouts. (b) Design of the nonlinear vibratory metamaterial with bi-linear bending stiffness in each unit cell, resembling the leaky ReLU activation function. (c) Readout strain locations of the metamaterial surface. (d) Time-series two-tone input voltage signal (k = 2) to excite the metamaterial. (e) Time-series readout strain signal of the metamaterial in response to the input. A subset of five signals are displayed. (f) Time-series output signals of an embodied task predicting a strain rate at a surface location, achieving \(R^2\) values of 0.91 and 0.84 on the training set and the test sets, respectively. (g) Time-series output signals of an embodiment-independent task of emulating ReLU-10 function, achieving \(R^2\) values of 0.90 and 0.85 on the training set and the test sets, respectively.}
    \label{fig:section 1}
\end{figure*}

We present a nonlinear mechanical metamaterial that functions as a physical reservoir, shown in Figure \ref{fig:section 1}. The metamaterial structure is 3D-printed using thermoplastic polyurethane (TPU) in a 5 by 5 network grid, where neighboring masses are connected through nonlinear beam elements (Fig. \ref{fig:section 1}-a). Contact nonlinearity arises from the bi-linear bending stiffness of each beam element, which relates curvature to bending moment. Under external excitation, the beam exhibits stiff bending when the triangular cutout gap closes, and compliant bending when the gap opens. This bi-linear stiffness property is intentionally designed in a leaky rectified linear unit (Leaky ReLU) form, resembling the activation function used in feedforward neural networks \cite{maas2013rectifier}, with the bending stiffness slope ratio governed by the geometry of the unit cell. The triangular cutouts are distributed across the metamaterial with randomly assigned orientations, facing either the \(+z\) or \(-z\) direction. An experimental characterization of the force-displacement behavior of a single ReLU unit cell is provided in SI.

In Physical Reservoir Computing (PRC), an input signal is applied to a nonlinear physical system that transforms it into a high-dimensional dynamic response. The resulting states are measured through readout signals, captured by sensors distributed across the system. The desired task output is then reconstructed using a linear combination of these readouts after training, allowing the physical dynamics of the system to process information and perform computation. Figure \ref{fig:section 1}-a shows the experimental setup used to validate the metamaterial as a physical reservoir. The metamaterial is clamped on two sides in \(+x\) and \(-x\) direction, and is excited by a linear modal shaker to generate an external force input signal acting on the metamaterial in the out-of-plane direction (\(z\)-direction), while the input force is measured using a load cell. The full-field deformation of the metamaterial is captured using high speed cameras and stereo Digital Image Correlation (DIC), enabling simultaneous time-resolved measurement of in-plane strain and 3-D motion across the entire structure.

A multi-tone input voltage signal with two frequency components, \(I_2(t) \), (shown in Equation~\ref{eq:Ik} where \(k = 2\)), is applied to the modal shaker to excite the ReLU metamaterial. The resulting strain field is recorded using stereo DIC, from which 78 virtual surface strain signals are extracted as readouts for PRC training, including the combination of \(\epsilon_{xx}\) and \(\epsilon_{yy}\) across the metamaterial surface. These readouts are trained using linear regression to perform a set of computational tasks. The strain readouts are trained for each task independently, enabling multitasking \cite{nakajima2020physical}. Once trained, the predicted signal is generated through a linear combination of the readout signals with their corresponding trained weights. Prediction accuracy is quantified using the coefficient of determination, or \( R^2 \), between the predicted and target signals. \( R^2 \) of 1 indicates a perfect fit between two signals. More details about the experimental method can be found in \hyperref[sec:setup]{Experimental Method}.

Figure \ref{fig:section 1}-c shows that the metamaterial can compute two different classes of tasks: embodied tasks and independent tasks. Here, we introduce embodied tasks as tasks that explicitly depends on the dynamics of the system itself. We perform proprioception and predict the strain rate signal, defined as the time-derivative of a strain at the location indicated in Figure \ref{fig:section 1}-c. Using all 78 readout strain sensors, the metamaterial achieves a performance of \(R^2_{train}=0.91\) and \(R^2_{test}=0.84\). An embodiment-independent task (i.e. an independent task) is a benchmark computational task that only depends on the external input to the system. As a representative independent task, we chose to emulate a ReLU-based benchmark function (shown in Equation \ref{eq:relu}) \cite{nelson2024spectral}, to synergize with the nonlinear properties of the ReLU-form bending stiffness in the metamaterial.

\begin{equation}
    y(t) =
    \begin{cases}
    10I(t), & \text{when } I(t) \ge 0, \\
    I(t), & \text{when } I(t) < 0 .
    \end{cases}
    \label{eq:relu}
\end{equation}

The ReLU-based task used throughout this study follows an equation in which the positive slope is ten times greater than the negative slope, and is therefore denoted as the ReLU-10 task. The metamaterial achieves a performance of \(R^2_{train} = 0.90\) on the training data and \(R^2_{test}=0.85\) on the test data. The system successfully predicts a strain rate signal at a selected location using only its strain signals (demonstrating an ability to take a time-derivative), while also performing an embodiment-independent task by emulating a ReLU activation function.

\section*{Importance of Nonlinearity for Computation}
\label{sec:NL vs L}
\begin{figure*}[tb!]
    \centering
    \includegraphics[width = 16 cm]{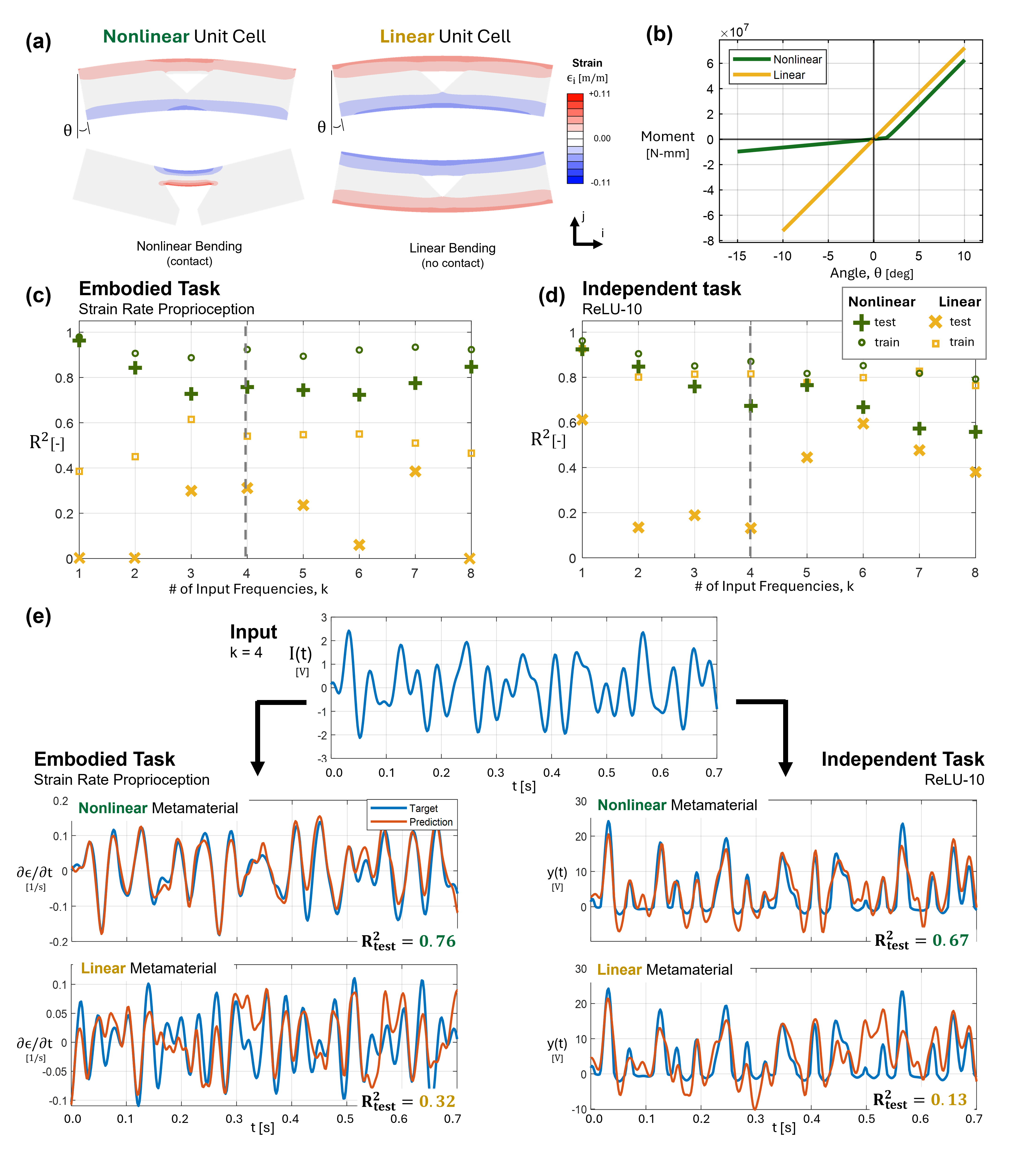}
    \caption{PRC performance of nonlinear and linear metamaterials under varying input complexities. (a) Schematic bending deformation of the nonlinear and linear unit cells, obtained from finite element analysis. (b) Moment-rotation bending stiffness of the unit cells used in the nonlinear and linear metamaterials, obtained from finite element analysis. (c) \(R^2\) values for the embodied task (strain-rate prediction) as input complexity increases. (d) \(R^2\) values for the structure-independent task (ReLU-10) as input complexity increases. (e) Time-series input under a four-tone signal excitation (k = 4, a scenario denoted as the dashed lines in panels -b and -c), and the corresponding embodied and independent task outputs for both nonlinear and linear metamaterials.} 
    \label{fig:section 2}
\end{figure*}

We demonstrate the importance of nonlinearity for a metamaterial's reservoir computing performance by comparing a nonlinear metamaterial with a linear version and  evaluating their performance across inputs of varying complexity. The linear version of the metamaterial is composed of unit cells with linearized stiffness where the linear unit cells have the same geometry as the nonlinear unit cells but do not have a cut and thus do not experience a contact nonlinearity, shown in Figure \ref{fig:section 2}-a. Finite element analysis was performed on individual nonlinear and linear unit cells to characterize their bending stiffness, as shown in Figure \ref{fig:section 2}-b. Details of the finite element analysis for both the nonlinear ReLU unit cell and the linearized unit cell are provided in the SI, Figure S1. The resulting moment-rotation curves show that the nonlinear unit cell (green curve) exhibits a bi-linear ReLU-like bending stiffness, whereas the linear unit cell (yellow curve) displays the same stiffness slopes between the two bending directions. For the nonlinear ReLU, the bi-linear switching between the stiffness regimes occurs at a nonzero angle of $\theta=2^\circ$; this backlash is due to the small initial gap in the unit cell between the contacting faces. The experimental characterization of the nonlinear ReLU unit cell is detailed in SI Figure S2. While this linearized metamaterial retains minor nonlinearities due to the asymmetric geometry of the triangular cutout, any nonlinearities in the linearized metamaterial are significantly weaker than the ReLU nonlinear metamaterial; this is clear in the stiffness comparison show in Figure \ref{fig:section 2}-b.

Fig. \ref{fig:section 2}-c and -d illustrates the robustness of PRC prediction performance in the nonlinear metamaterial across increasing input complexity with results from the linear counterpart overlaid for comparison, evaluated using the same two tasks: an embodied task, which involves predicting the strain rate at the same surface location, presented in Figure \ref{fig:section 1}, and an independent task, namely the ReLU-10 task. Experiments are conducted under identical input conditions to ensure a fair comparison between the two systems. The different input conditions are varying input complexity, defined as the number of frequency components embedded in the signal, detailed in \hyperref[sec:Input Signals]{Input Signals} in Methodology Section. The nonlinear metamaterial consistently outperforms the linear metamaterial across the range of input complexities for both of the tasks. As input complexity increases, the nonlinear system exhibits only a slight reduction in predictive accuracy for both tasks, indicating its robustness in task difficulty. This slight degradation in performance observed for the nonlinear system is expected. This is because the corresponding tasks become more difficult to predict as the input becomes more complex with additional frequency components. In contrast, the linear metamaterial displays irregular and largely unstructured performance, with no clear dependence on input complexity, particularly in the test sets where \(k = 5, 6\). This phenomenon suggests that the linear system lacks generalization ability and exhibits high sensitivity to inputs, as evidenced by the large dependency in \(R^2\) values between the training sets and the test sets, resulting in overfitting. Moreover, linear systems tend to produce more linearly dependent information between the readouts, resulting in random and inconsistent performance \cite{he2025role}.

An example of the time-series performance under a four-tone input excitation is shown in Figure \ref{fig:section 2}-e for both the embodied task and the independent task. When performing the embodied tasks, the nonlinear system outperforms the linearized system. It is noted that the target output signals computed are different between the two systems although it is the same task. In both cases, the task is to predict the strain rate at the same spatial location as shown in Figure \ref{fig:section 1} (i.e. proprioception), and strain rate depends on the global dynamics of the structure. Since the linear and nonlinear metamaterials exhibits different dynamics due to slight different overall stiffness, they produce different target signals. This highlights a key feature of the embodied task: the task inherently depends on the dynamics of the system. In contrast, the independent task (ReLU-10) depends only on the input, and since both the nonlinear and linear metamaterial are excited under identical input conditions, the two systems are evaluated against the same target signal. This result indicates that the nonlinear metamaterial outperforms the linear metamaterial for independent tasks with identical output signals and embodied tasks with different output signals.

\section*{Information Separation Mechanism}
\label{sec:PCA}

To understand the underlying mechanism by which the nonlinear metamaterial processes information, we analyze how frequency content is distributed across the reservoir readout sensors. We first analyze the spectral responses of the metamaterial's readouts and then apply Principal Component Analysis (PCA) \cite{abdi2010principal} to identify the dominant dimensions of the frequencies distributed across these readouts. These dominant dimensions of the system's readout frequencies reveal the main dynamical pattern and provides an understanding of how the information is processed within the system's dynamics. We relate the sensor distribution in the principal-component space to physical quantities, including signal-to-noise ratio, nonlinearity, and correlation. Additional details of the frequency analysis and signal-to-noise ratio are provided in the \hyperref[sec:Methodology]{Methodology}.

\begin{figure*}[tb!]
    \centering
    \includegraphics[width=\textwidth]{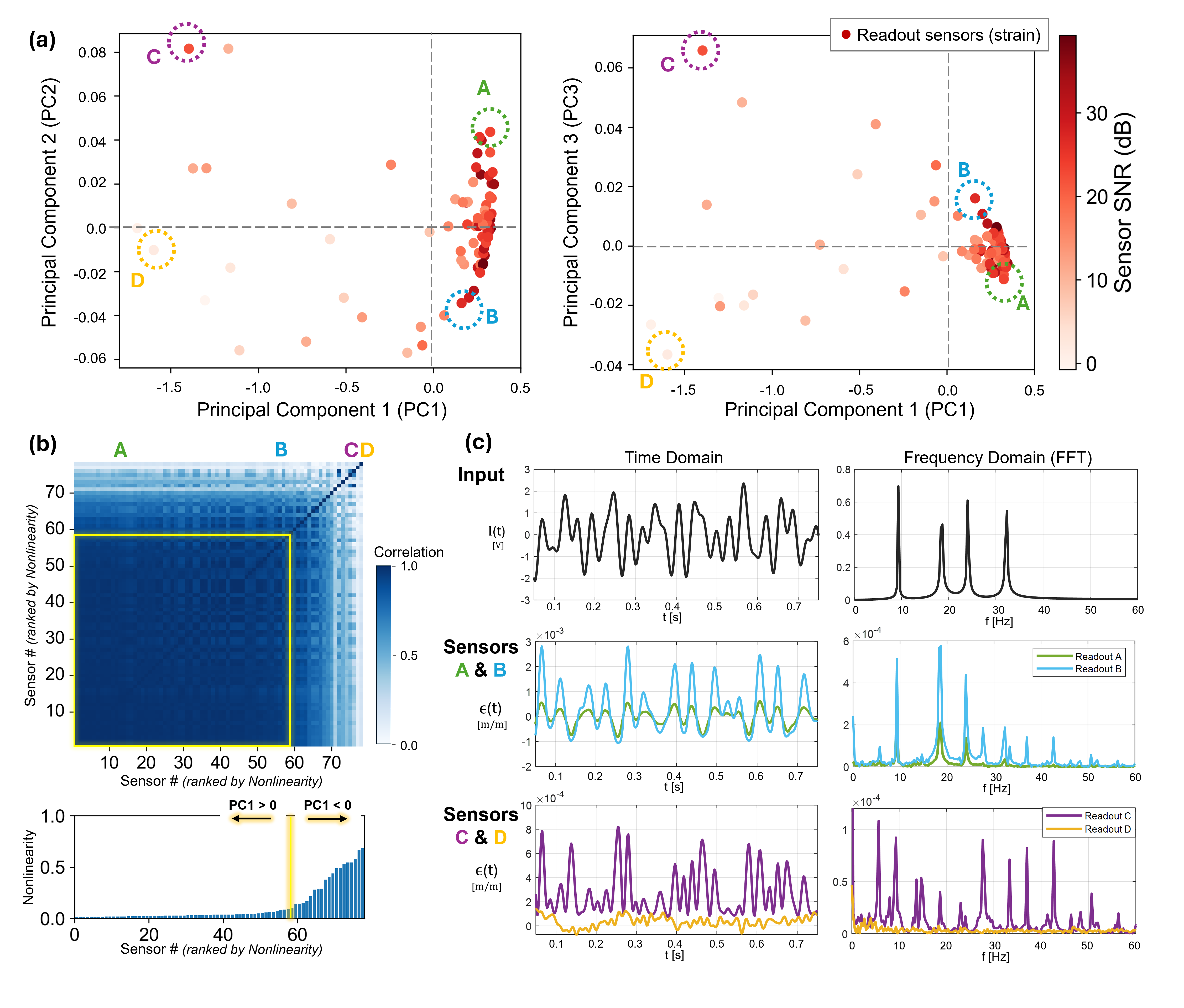}
    \caption{Principal Components Analysis of the frequency content of the reservoir's readout signals to explain the frequency separation mechanism. (a) representation of the first 3 major principal components that exhibit 90.5 \% of the variance. Each dot represents a sensor signal with their color indicating the sensor's signal-to-noise ratio. The readouts are more distributed along PC-2 while cluster within PC-1 and PC-3 with several outliers. (b) Correlations of all 78 readout signals ranked by nonlinearity with the four selected representative readouts labeled in the nonlinearity metric order. First ranked 59 sensors within the yellow box and line indicate the readouts that cluster in positive PC-1 space. (c) time-series and frequency content of the input signal and four selected readout signals: A (green) and B (blue) show the two far ends within the cluster along PC-2; C (purple) shows a high SNR outlier readout sensor; D (yellow) shows a low SNR outlier readout.}
    \label{fig:section 3}
\end{figure*}

The original 600-dimensional frequency-domain representation of the metamaterial reservoir readouts is reduced to three major principal components (PCs) that explains 90.5\% of the total variance. Figure \ref{fig:section 3}-a visualizes all 78 readouts in the major 3-dimensional principal component space, projected onto two 2-D planes (PC-1 vs. PC-2, and PC-1 vs. PC-3), indicating that the metamaterial reservoir dynamics are effectively low-dimensional. Each dot in the plots represents an individual readout signal. In the PC space, it is observed that some readout sensors cluster together while others are more widely spread out. This distinction is strongly related to signal quality with high SNR readout signals are largely concentrated within the main cluster, whereas most low SNR signals correspond to the dispersed outliers. This association of the SNR of the readouts and PC cluster suggests that the majority of the distant readouts in PC-space arise from noisy measurements rather than significant information separated from the system's nonlinearity. In the main cluster, the PCA of the readouts' frequency content reveals a distinct strip-like distribution along the PC-2 direction, while these readouts remain clustered in PC-1 and PC-3. This indicates the frequency content associated with PC-2 exhibits greater information separation between readout sensors, whereas the dominant frequency content captured by PC-1 and PC-3 remain consistent.

We quantify the nonlinearity of each sensor and evaluate the correlation between the sensors' information. Figure \ref{fig:section 3}-b presents the pair-wise correlation matrix of all 78 readouts. A value of 1 indicates perfect correlation between two readout signals, and 0 indicates no correlation. In Figure \ref{fig:section 3}-b, the sensors are ranked by their nonlinearity. The nonlinearity metric captures that amount of additional frequency content that is in a sensor vs. the frequencies that were input into the system. The nonlinearity ranges from 0 and 1 where 0 indicates a linear sensor that has the same frequency content as the input and 1 indicates a highly nonlinear sensor with significant frequency content that was not in the input. The detailed description of how to quantify nonlinearity is shown in \hyperref[sec:Methodology]{Methodology} and is further used in \hyperref[sec5:NL_MM]{Multi-tasking of Metamaterial Reservoir} with a memory metric to quantify a series of tasks. Readouts with low nonlinearity (bottom-left region of the matrix) exhibit high correlation, indicating strong mutual information among them. These are also the readouts that are clustered showed in PC-space. As nonlinearity increases, correlation diminishes, and signals become more distinct (upper-right corner of the matrix). The uncorrelated readouts are the sensors that scatter as outliers in the PC-space. PC-1 roughly captures the variation in nonlinearity across the readouts, as highlighted by the yellow box in the correlation matrix and the dividing line in the nonlinearity plot in Figure~\ref{fig:section 3}-b, which separates the readouts into regions corresponding with PC-1 \(<\)0 and PC-1 \(>\)0. This result demonstrates that the readout sensors clustered in the PC space exhibit higher correlation with each other and low nonlinearity (PC-1 \(>\) 0), and those that are spread as outliers are corresponding to low correlation with high nonlinearity (PC-1 \(<\) 0).

To further analyze the distribution of the readout information in the PC-space, in particular the strip-like spread along the PC2 direction and the outlier sensors, we examine several selected representative readouts from distinct regions of the PC plot, labeled as A-D in Figure \ref{fig:section 3}-a and -b. Figure \ref{fig:section 3}-c shows the time-series signals and the corresponding fast Fourier transforms (FFT) of the 4 selected readouts, with the input signal (black curve) as a baseline reference. Readouts A and B (green and blue labels, respectively) lie within the cluster region but on the two opposite sides of PC-2. The FFTs of these readouts indicate that PC-2 corresponds to whether lower or higher frequencies are emphasized. Two additional sensors were selected from the outlier region, representing one high-SNR case and one low-SNR case. Readout C (purple label), an outlier in the PC-space but with high SNR, exhibits multiple additional frequency components beyond the primary frequencies from the input. In contrast, another outlier readout with low SNR, labeled as Readout D, displays low amplitude and irregular fluctuation in both time and frequency domain, further confirming it is a noisy signal with no significant dynamic information.

These findings show that the separation of frequency content within the reservoir readout states is the governing mechanism that enables information processing in a mechanical reservoir computer. The dimensionality analysis (PCA) reveals that the system’s dynamics are reduced down to three primary dynamic modes that capture 90\% of the dynamic complexity, each captured by one of the principal components. Readouts that cluster in PC space correspond to high-SNR signals that exhibit strong mutual correlation, indicating meaningful dynamical information. In contrast, readouts scattered away from the main cluster are predominantly noisy information, characterized by their low-SNR. Further, sensors that are spread along the PC-2 direction capture different types of frequency content. The combined analysis of PCA, SNR, nonlinearity metric, and correlation provides a framework to identify meaningful nonlinear information separation and distinguish from measurement noise in mechanical reservoir computing. Further details of the PCA analysis and comparison between the nonlinear and linear metamaterials are shown in SI Appendix Figure S3.

\section*{ The Role of Sensor Selection}
\label{sec:sensor selection}

PRC performance not only depends on the nonlinear dynamics of the system, but also on which readout sensors are used to capture those dynamics. Up to this point, PRC performance has been evaluated using a full set of 78 readout sensors. The dimensionality analysis in \hyperref[sec:PCA]{Information Separation Mechanism} reveals how frequency content is separated in the metamaterial's response and shows the majority of the dynamics can be explained using only three dimensions. This observation naturally raises the question of whether the same frequency-based analysis can be leveraged to identify a minimal yet sufficient set of readouts to perform a specific task. To understand the role of frequency content in a sensors usefulness for the task,  we use a greedy sensor selection algorithm with the frequency content alignment metric described in SI to perform sensor selection on the embodied task of predicting a strain rate. This enables intelligent selection of the most task-relevant readout sensors based on their frequency compatibility to a given task. Beginning with the sensor with the greatest frequency content alignment with the task, we iteratively add the sensor that produces the best total frequency content alignment with the task when combined with the subset of selected sensors. This process is repeated until all 78 sensors are ranked (The full set of all 78 readouts locations is shown in Figure \ref{fig:section 1}-c). Finally, the number of sensors that yields the highest prediction accuracy is selected. As a baseline, we also perform naïve sensor selection by randomly choosing the same number of sensors from the full set of 78 readouts and repeating this process 1000 times.

\begin{figure*}[tb!]
    \centering
    \includegraphics[width=\textwidth]{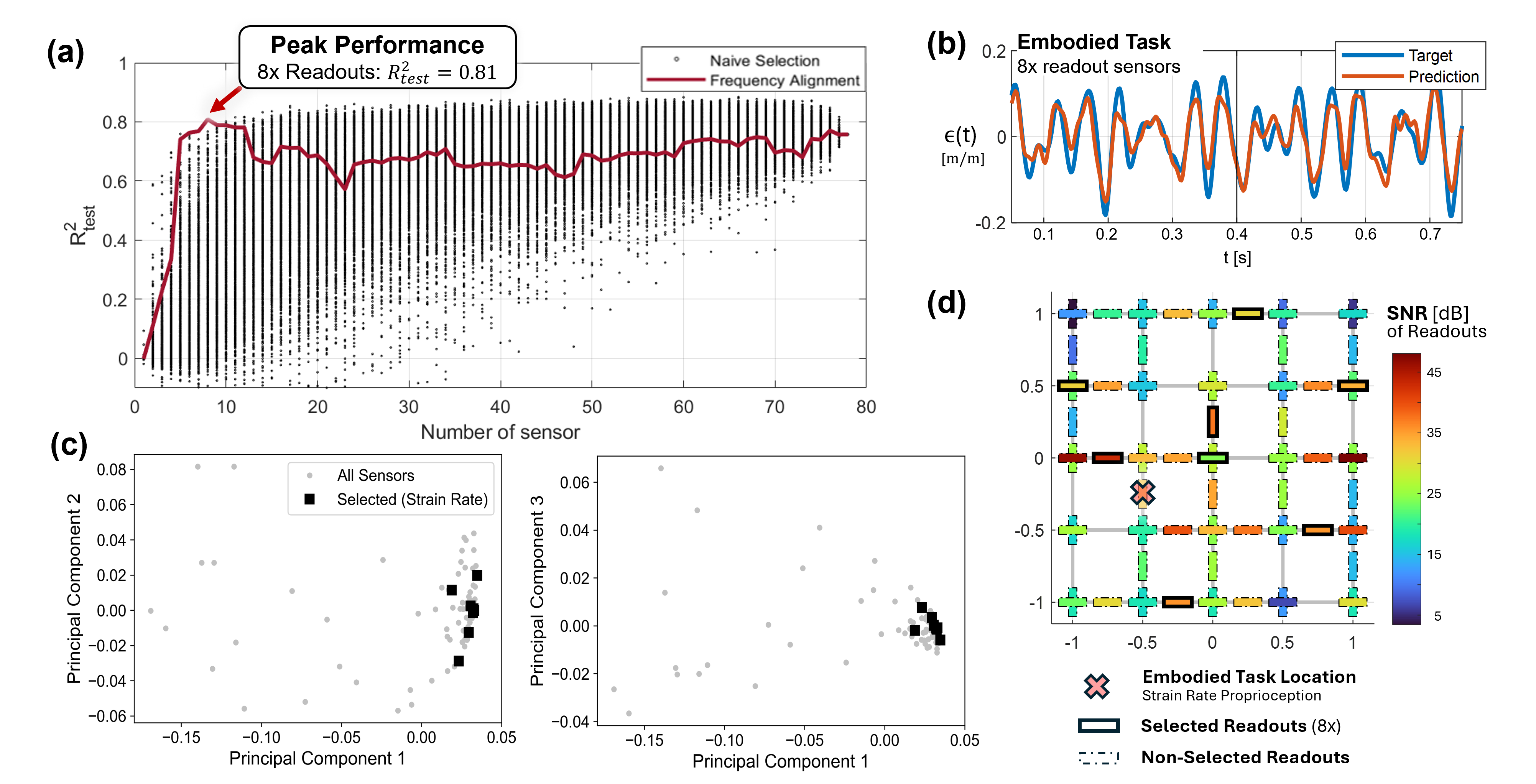}
    \caption{Prediction performance as a function of readout sensors under a four-tone input case. (a) \(R^2\) values for the strain-rate prediction task as sensors are incrementally added using a frequency alignment method (red solid line). For comparison, performance obtained from naïve sensor selection is shown using 1000 different random sensor combinations (black dots). Only eight sensors are used to achieve the highest \(R^2\) value of 0.81 compared to the complete set of 78 sensors with \(R^2\) value of 0.76. (b) Output signals of the embodied task using the 8 selected readout sensors, for the embodied task. (c) The selected 8 readout sensors highlighted in the PC-space, for the embodied task. (d) Physical locations of the readout sensors with their signal-to-noise ratio in dB. The eight selected sensor locations for the embodied tasks are indicated using thick solid marker line. Horizontal rectangular markers denote \(\epsilon_{xx}\) and vertical markers denote \(\epsilon_{yy}\).}
    \label{fig:sensor selection}
\end{figure*}

Figure \ref{fig:sensor selection}-a shows the PRC performance for the embodied strain rate task using both the frequency-based sensor selection algorithm and naïve selection. The solid curve represents the performance on the test set, \(R^2_{test}\), obtained by incrementally adding sensors using the greedy selection algorithm. Naïve sensor selection is used for comparative purposes (shown as black dots) for the same number of sensors. The frequency alignment method using greedy selection algorithm converges to peak performance using only eight sensors, achieving \(R^2 = 0.81\) and outperforming the other 1000 combinations of randomly selected sensors. In contrast, using all 78 sensors degrades performance to \(R^2 = 0.76\). The time-series task performance using the reduced set of readouts is shown in Figure \ref{fig:sensor selection}-b. To understand why specific sensors are selected for the embodied task, the chosen readouts are highlighted in the principal component (PC) space, shown in Figure \ref{fig:sensor selection}-c. The selected sensors lie in the PC-1/PC-3 cluster and exhibit variation along the PC-2 direction, where low- and high-frequency content are distributed. This distribution suggests that successful task performance requires a combination of frequency components across the spectrum and those align well with the tasks. A closer examination of the physical locations of the selected sensors, shown in Figure \ref{fig:sensor selection}-d, reveals that most of them are \(\epsilon_{xx}\) strains, which exhibits consistently higher SNR compared to the \(\epsilon_{yy}\) strains. This difference in SNR is due to the boundary conditions of the metamaterial, which is clamped on the horizontal boundaries and free in the vertical boundaries. Under the out-of-plane input excitation, this boundary condition configuration induces a stronger bending strain along the x-direction, resulting in higher-SNR in \(\epsilon_{xx}\) measurements. The SNR is lower near the corners of the metamaterial, close to the mounts, where structural deformation is minimal. As a result, the sensor selection algorithm chooses high-SNR strain signals, as they carry the most significant dynamic information for task prediction.

The frequency alignment method ranks the readout sensors that contain task-relevant dynamics yielding the smallest number of sensors necessary to effectively recreate the relevant task. Adding more sensors beyond this optimal subset can degrade the performance, due to increased linear dependency and redundant information among the readouts \cite{he2025role}. Importantly, different tasks require different subsets of optimal sensors. This is because the frequency-alignment method is task-dependent, ranking each sensor based on their frequency compatibility with the given task. The SI Figure S4 presents the sensor selection for the ReLU task. While random sensor selection occasionally achieves comparable or better performance, the frequency alignment method with a greedy sensor selection algorithm consistently outperforms the vast majority of random sensor combinations for a small numbers of sensors. Together, these results demonstrates the importance of choosing sensors whose frequency content aligns with the task. This frequency alignment method enables efficient sensor reduction by avoiding an exhausting random sensor search, and improves prediction performance by eliminating any unnecessary noisy or linearly dependent information.

\section*{Multi-tasking of Metamaterial Reservoir}
\label{sec5:NL_MM}

Nonlinearity and memory are fundamental properties that determine a reservoir's ability to compute a task, as successful prediction requires alignment between the nonlinearity and memory of the reservoir's dynamics and those required by the tasks. We calculated the nonlinearity and memory of each readout sensor and of each task to examine their overlap in the nonlinearity-memory space. In reservoir computing, the readouts represent the observable states of the reservoir used to reconstruct the target tasks; therefore, the dynamics of these readouts determine which tasks the reservoir can effectively perform. In this section, we evaluate the performance across a range of tasks that vary in nonlinearity and memory with the tasks classified as either embodied or independent. The nonlinearity metric (Equation \ref{eq:nonlinearity}) used in this section was introduced in previous section (Information Separation Mechanism in Metamaterial PRC).

The memory requirement of each task was quantified by assessing the correlation of the task with time delays of the input signal. Since the input signal is not an independent and identically distributed (IID) signal, we cannot use the standard definition of the memory capacity \cite{jaeger2001short} to calculate the memory. Instead, we focus on identifying the time delay that produces the maximal correlation. The memory is quantified as the time-delayed correlation between the signal of interest and the input signal capturing both the time delay of peak correlation and the magnitude of the correlation. Memory is a value between 0 and 25 where 0 indicates no correlation and those no memory of past inputs and 25 indicates perfect correlation back 25 time-steps. In this study, the signals are sampled at 500 Hz, so 25 time-steps correspond to 0.05 second in real time. Tasks with large memory are strongly correlated with the input at a significant time delay. Tasks with low memory either exhibit low optimal time delay or low correlation with the time delayed input signal. Details of how memory delay is quantified are provided in \hyperref[sec:Methodology]{Methodology}.

\begin{figure*}[tb!]
    \centering
    \includegraphics[width=\textwidth]{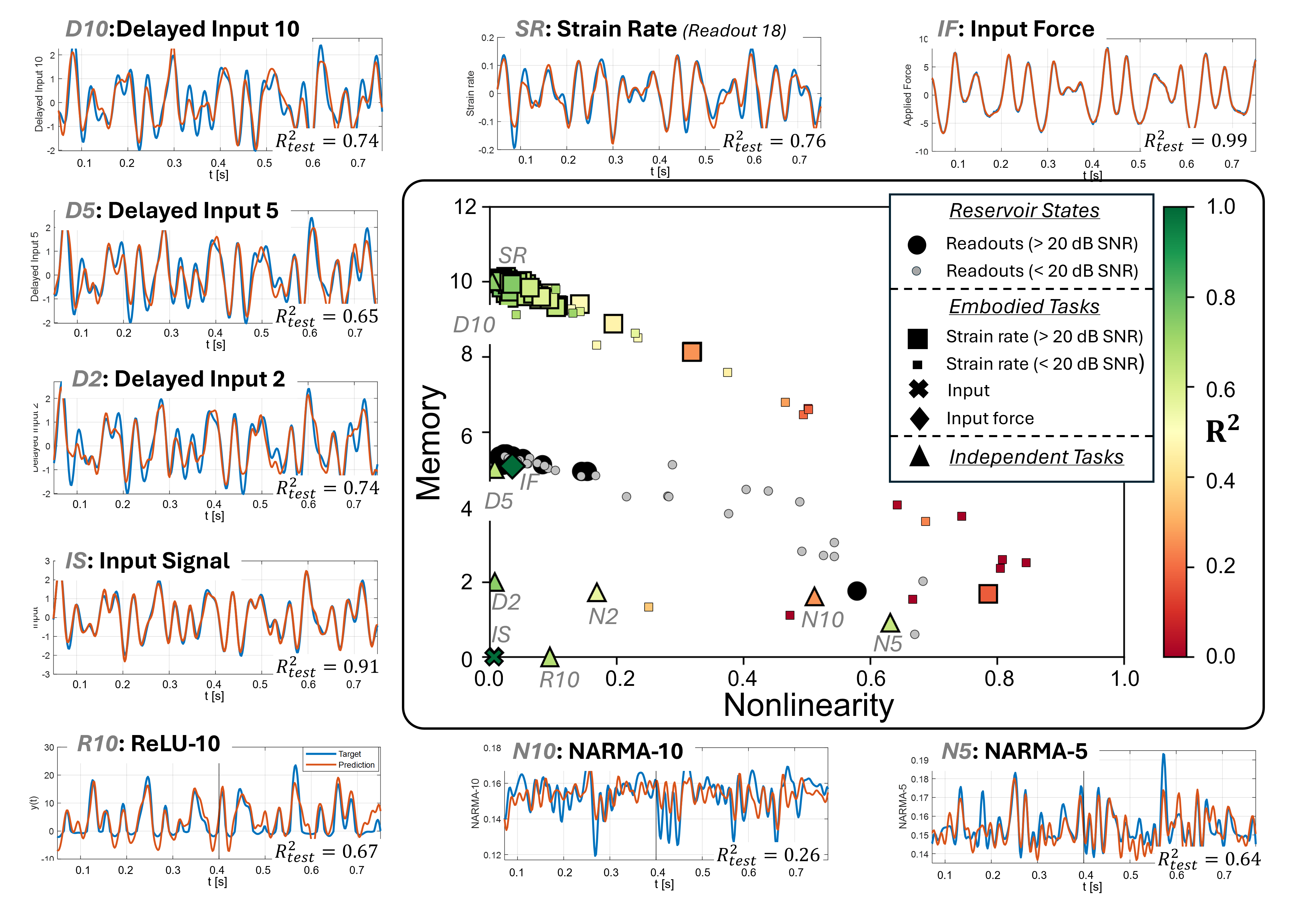}
    \caption{Mapping of the reservoir dynamics and the computational tasks in the nonlinearity vs. memory space, showing alignment of the reservoir state with the tasks. Prediction performance (\(R^2\)) of each tasks is denoted by color. All 78 sensors are quantified in a memory-nonlinearity metric space. The bigger marker size in the (strain) sensors and the (strain rate) embodied tasks indicate SNR above 20 dB as a threshold, while the smaller markers indicate SNR below 20 dB within these signals. Input signal is located on the origin, (0,0), of the metric space, labeled as "IS". Time-series output signals of the tasks prediction are shown as insets: SR = strain rate; IF = input force; R10 = ReLU-10; D2, D5, and D10 denote as Memory with 2, 5, and 10 time steps, respectively; N5 and N10 denote as NARMA-5 and 10, respectively. The \(R^2\) values of each of the task are shown within their corresponding inset plot.}
    \label{fig:NL_memory}
\end{figure*}

Figure \ref{fig:NL_memory} shows the distribution of all 78 strain readout signals and the prediction performance of the metamaterial reservoir across a range of computational tasks in the nonlinearity-memory metric space. In Figure \ref{fig:NL_memory}, all cases are evaluated under a four-tone input voltage signal. To capture the dynamics of the metamaterial reservoir, the nonlinearity and memory of the individual readout sensors are shown by the circular markers where sensors with low SNR are indicated in gray. To capture the dynamics of the tasks, the nonlinearity and memory of both embodied and independent tasks are shown. Embodied tasks, shown as square markers, include the measured input force and computation of each of the 78 strain-rate signals. Independent tasks, shown as triangular markers, include NARMA-5 and -10, memory-delay task of 2, 5, 10 steps, and the ReLU-10 task. PRC predictive performance for computing each task is represented by color according to their \(R^2\) values. Signal quality for all strains and strain-rate signals is distinguished by marker sizes. Larger markers represent signals with SNR greater than 20 dB, while smaller markers indicate SNR values below this threshold.

The 78 readout sensors (black and gray circular markers) are located in a distribution in the metric space, indicating they are diverse in terms of their nonlinearity and memory. The fact that these readouts are sparse and distributed demonstrate that the metamaterial reservoir exhibits nonlinear separation through the dynamical response. However, a decent amount of readouts are clusters in the low-nonlinearity mid-memory regions of the metric space, where the signals generally have high signal-to-noise ratio (SNR \(\geq\) 20 dB) and nonlinearity values below $\approx0.2$. In contrast, a handful of readouts display large nonlinearity, and they are generally associated with low SNR. This is consistent with the observation discussed in previous section \hyperref[sec:PCA]{Information Separation Mechanism}. The only high-SNR sensor located near a nonlinearity value of approximately 0.6 is Sensor C, shown in Figure \ref{fig:section 3}.

With the readout signals established in the nonlinearity-memory space, we next examine a series of task performance to determine how the reservoir's dynamics align with different computational demands in a unified framework. The input voltage signal is theoretically defined as the origin, (0,0), of the nonlinearity-memory space (cross marker), reflecting its low nonlinearity and low memory. As a result, predicting the input voltage signal is generally straightforward using the PRC method. The input force prediction (diamond marker) achieve the highest PRC performance among all the tasks. This strong performance is due to the fact that the force signal lies within the cluster of the readouts, indicating it has a similar dynamic information as the readout signals. As a result, the force signal can be effectively recreated using the linear summation of the readout states.

The strain rate signals (square markers), which are time-derivatives of strain readouts, are used here as embodied tasks. Small squares correspond with those strain rate signals with low SNR. The strain rate signals exhibit a similar pattern as the strains: high SNR signals tend to cluster in the low-nonlinearity region and low SNR signals mostly lie on high-nonlinearity. The reservoir generally predicts strain rate signals with high SNR well. This suggests that the reservoir is most effective for tasks that lie within a similar nonlinearity range as the reservoir readouts, typically nonlinearity less than 0.2. Because the nonlinearity metric is based on frequency content, this overlap indicates that the reservoir contains the spectral features needed to represent and reconstruct those tasks. The reservoir is unable to predict two of the strain-rate tasks associated with high SNR because the time derivative increases the effective nonlinearity of the signals. The memory of the strain rate tasks is higher than that of the corresponding readout strains because taking the time derivative introduces a phase offset, thus increases the memory. Memory does not appear to have a strong influence on prediction performance.

Another observation is that there is a general tradeoff of the reservoir's readouts as well as the strain rate tasks between nonlinearity and memory where these signals either have high nonlinearity or high memory. Although the metrics are calculated differently in this study, this trend is qualitatively consistent with the known nonlinearity-memory tradeoff in reservoir computing literature \cite{dambre2012information, inubushi2017reservoir}. The result highlights the important role of signal quality interpreting nonlinearity within the reservoir readouts. Many readouts exhibit high nonlinearity but low SNR, indicating that the observed nonlinear frequency content arise from noise rather than structured dynamics. Because noise introduces broadband frequency components that are not present in the input, it can appear as strong nonlinearity in the frequency analysis. Therefore, the reservoir has difficulty predicting such high-nonlinearity yet low-SNR tasks.

Evaluating a range of independent benchmark tasks enables comparison with existing reservoir computing studies and helps quantify the generality of the reservoir's computational capability. We next evaluate a series of independent tasks, including the ReLU-10, input-delay tasks with 2, 5, and 10 time step delays, and NARMA-5 and -10, all shown in triangular markers in Figure \ref{fig:NL_memory}. Tasks that remain in the low-nonlinearity region, despite high-memory, generally achieve good prediction performance, as demonstrated by the memory delay tasks and the ReLU-10. In contrast, tasks located in the high-nonlinearity region tend to exhibit degraded performance, most notably NARMA-10. This is because successful task prediction typically requires multiple informative readouts located in proximity to the task within the metric space. Since the readouts near this task are dominated by low-SNR, the resulting prediction performance is expected to be low.

Although NARMA-5 and NARMA-10 appear adjacent in the nonlinearity-memory space, their prediction performance differs. This is because the nonlinearity metric captures the overall presence of nonlinear frequency content but does not distinguish which specific nonlinear frequencies are generated. Therefore, two tasks may locate in a similar nonlinear-memory space but do not necessarily require the same frequency content from the readouts. In this particular system, the nonlinear frequency components produced by the metamaterial reservoir appear to align more closely to those in NARMA-5, than those in NARMA-10. As a result, the available readouts provide sufficient information to reconstruct NARMA-5, but fail to perform for NARMA-10.

Mapping both reservoir readouts and computational tasks in the nonlinearity-memory metric space provides a useful framework for qualitatively assessing the compatibilities between a reservoir and the task. The framework highlights that only readouts with both good signal quality and sufficient nonlinear dynamics contribute meaningful computation for PRC. In particular, sensors with low SNR appear highly nonlinear and lack correlation with the input and therefore provide little useful information for prediction. The nonlinearity-memory mapping offers insight into the designs of mechanical metamaterial reservoir. Tasks located in highly nonlinear regions of the space requires reservoirs capable of generating stronger nonlinear response. Therefore, designing metamaterial with high nonlinearity enables improved performance on more difficult tasks such as high-order NARMAs.

\section*{Conclusion}
In this study, we established the information processing capability of a metamaterial reservoir computer. The nonlinearity arises from the bi-linear ReLU bending stiffness of each spring-like element. We experimentally validated its information processing capabilities by performing an embodied task and an independent benchmark task under external vibratory excitations. We evaluated the PRC performance of the metamaterial across varying input complexity and compared it against a linear metamaterial counterpart. Our results showed that the nonlinear metamaterial successfully performs these two tasks with high accuracy and is robust across various input complexity, outperforming its linear counterpart.

We investigated the mechanism of information separation by analyzing the frequency content of the readout signals using Principal Component Analysis (PCA), a dimensionality-reduction technique. PCA reveals the fundamental mechanism of computation in the mechanical reservoir: the nonlinear dynamics of the metamaterial transforms input frequencies into a broadened spectrum of frequency components in the system's response. The diverse spectral features in the readout signals provide meaningful dynamics information, enabling efficient PRC performance. We identified the most informative readouts by quantifying the compatibility between each readout and a given task using a frequency alignment method. The approach significantly reduces the number of required sensors without compromising performance and consistently outperforms randomly selected sensor combinations. Successful prediction requires the frequency content of the reservoir to align with that of the task, enabling the selection of the most informative sensors and facilitating sensor reduction.

Finally, we demonstrated the multi-tasking capability of the metamaterial by evaluating the computational performance on a series of embodied and independent tasks. Together with the readout signals, these tasks are quantified within a unified nonlinearity-memory metric space. Mapping both the reservoir dynamics and the tasks in the nonlinearity-memory metric space provides a framework to evaluate reservoir performance and guide reservoir design.

This work establishes a foundational framework for designing mechanical metamaterials that function as physical reservoirs. We demonstrated that a nonlinear mechanical metamaterial can not only sense its own dynamic behavior but also compute a range of computational tasks, providing a framework to harness the embodied intelligence of any mechanical metamaterial. This study enables the design and development of adaptive structures that sense, compute, and respond through their own physical behavior and opens new opportunities for embodied intelligence in soft robotics, morphing aerospace structures, and other engineered systems.

\section*{Methodology}
\label{sec:Methodology}

\subsection*{Experimental Method} 
\label{sec:setup}

We conduct experiments to validate the mechanical metamaterial’s capability to process information using the PRC method. We used the modal shaker actuator (Baldor Motors and Drives, Serial\#: 200618) with a prescribed time-varying input signal to the metamaterial in an out-of-plane direction. The metamaterials are clamped on an 80-20 mounting structure on their left and right sides as a boundary condition, but the top and bottom sides are free. To measure the time-varying input forces produced by the modal shaker, we attached a 6-Degree-of-freedom load cell transducer (ATI Industrial Automation, Nano 17) between the shaker and the metamaterial. We capture the in-plane strain and 3D deformation of the entire surface of the metamaterial using stereo Digital Image Correlation (Stereo DIC). To perform Stereo DIC setup, we focused two high speed cameras (Phantom VEO440L) on the surface of the metamaterial with a small angular offset between the two cameras. We applied a black-and-white speckle pattern on the surface of the metamaterial to facilitate image tracking, and we illuminated the setup using high power LEDs. We used Correlated Solutions Vic-3D software to extract surface strains from the time-varying image pairs. We ran 8 different test cases with different input complexity from a single sine wave to 8-tone wave signal. The frequencies of the multi-tone signals were randomly chosen between 5 Hz - 50 Hz. The DIC images were captured at a sample rate of 500 frames per second, and at a resolution of 1920 $\times$ 1080 pixels for all test cases. All time-varying signals are time-synced during testing using NI-DAQ 6363. All time-synced signals are recorded for 6 seconds, yielding a total of 3000 time steps per signal. To focus on only the steady-state dynamics for PRC performance, we discard the first 30\%\ and last 20\%\ of the recorded data. This pre-processing results in 1500 samples per dataset. The remaining data are partitioned into training and testing sets, with the first 80\%\ used for training and the final 20\%\ for testing.

\subsection*{Input Signals}
\label{sec:Input Signals}
The input signals are constructed as a multi-tone signals as defined in Equations \ref{eq:Ik_hat}, \ref{eq:Ik}, where increasing the number of frequencies corresponds to higher input complexity. A complexity value of $k=1$ corresponds to a single-frequency sine input, while a value of $k=8$ corresponds to a multi-tone signal formed by the superposition of eight sines. The frequencies for the input signal are randomly selected within a frequency window of \([0,40] Hz\) where the natural frequencies of the metamaterial fall within. The sequence of input frequencies as $k$ increases from 1 to 8 is given as \(f_k = [9.3, 24.1, 32.2, 18.5, 38.0, 14.5, 21.0, 11.1] Hz\). In Equation \ref{eq:Ik_hat}, \(RMS\) is root-mean-square of the signal, defined as the square root of the mean of the squares values over time.

\begin{equation}
    \hat{I}_k(t)=I_k(t)/RMS(I_k(t))
    \label{eq:Ik_hat}
\end{equation}

\begin{equation}
    I_k(t) = \sum_{i=1}^{k} \sin\left( 2\pi f_i t \right), (k = 1,2,\ldots,8)
    \label{eq:Ik}
\end{equation}

\subsection*{Frequency Content Analysis}

Kiyabu et al. \cite{kiyabu2025optomechanical} developed a frequency content analysis that characterizes a reservoir's ability to produce different frequencies with arbitrary phase shift and independence from other frequencies. To calculate the frequency content ($\chi$), the reservoir's state trajectory matrix $\mathbf{X}$ is transformed into a basis of cosines and sines via a discrete Fourier transform, yielding matrix $\mathbf{X}_f$. Trajectories composed of $2n$ time steps in the time domain will transform into $2n$ bases in the frequency domain: $n$ cosine bases and $n$ sine bases for $n=600$ discrete frequencies. Within this high dimensional frequency space, the cosine and sine basis vectors are projected onto the subspace formed by the transformed state trajectories (i.e., the range of $\mathbf{X}_f$). For frequency $i$, the frequency content is defined as: 
\begin{equation}
    \chi_i = \frac{1}{2}(||\mathbf{P}\mathbf{u}_i||_2^2 + ||\mathbf{P}\mathbf{u}_{i+n}||_2^2)
\end{equation} 
where $\mathbf{P}$ is the projection matrix onto the transformed state subspace in the frequency domain, $\mathbf{u}_i$ is the unit basis vector corresponding to the cosine of frequency $i$ and $\mathbf{u}_{i+n}$ is the unit basis vector corresponding to the sine of frequency $i$. If $\mathbf{X}_f=\mathbf{U_f\Sigma_f V_f}^T$ is the singular value decomposition of the transformed matrix $X_f$, then the projection matrix can be calculated as $\mathbf{P}=\mathbf{U_f} \mathbf{U_f}^T$. In practice, if the rank of the state matrix is $k$, the first $k$ columns of $\mathbf{U_f}$ are included, i.e., $\mathbf{U_f} \in \mathcal{R}^{2n \times k}$. To account for noise in the experimental strain readouts, the effective rank $k$ was calculated as the number of principal components required to explain 99.99\% of the variance of the reservoir's state matrix. $\chi_i$ ranges from 0 to 1 and indicates how well the reservoir can produce outputs with frequency $i$, where 0 indicates no ability to produce frequency $i$ and 1 indicates perfect ability to produce frequency $i$ with arbitrary phase and amplitude independent of other frequencies. Kiyabu et al. \cite{kiyabu2025optomechanical} also introduced a measure of frequency alignment with a given task: 
\begin{equation} \label{eq:frequency_content_alignment}
    \hat{\chi}_{task}= \frac{\sum_{i=1}^n c_i^2 \chi_i}{\sum_{i=1}^n c_i^2}
\end{equation}
where $c_i$ is the discrete Fourier transform coefficient of the computational task at frequency $i$. $\hat{\chi}_{task}$ possesses a value between 0 and 1 and corresponds to the expected value of the coefficient of determination, i.e., $\hat{\chi}_{task} = \mathbb{E}[R^2]$, over all possible tasks with the given set of $\{c_i\}$. In order to characterize the individual readouts, the frequency content was calculated for each of the 78 sensors individually. The analysis was performed on individual columns of $\mathbf{X}$ rather than the entire state matrix.

\subsection*{Signal-to-Noise Ratio}
We estimate the signal-to-noise ratio for each sensor. If $\mathbf{X}=\mathbf{U\Sigma V}^T$ is the singular value decomposition of the reservoir's state matrix, a truncated reconstruction is produced via $\mathbf{X}_k=\mathbf{U_k\Sigma_k V_k}^T$, where only the first $k$ singular values and vectors are retained. The effective rank $k$ was previously calculated during the frequency content analysis. The remainder of the information in the state matrix was considered noise: $\mathbf{X}_{noise} = \mathbf{X} - \mathbf{X}_k$. The 78 columns of $\mathbf{X}_k$ were compared against the 78 columns of $\mathbf{X}_{noise}$ to determine the signal-to-noise ratio in decibels for each sensor: 
$SNR = 10 \log(\frac{\langle x_k^2 \rangle}{\langle x_{noise}^2 \rangle})$, where $x_k$ is a column of $\mathbf{X}_k$, $x_{noise}$ is a column of $\mathbf{X}_{noise}$, and $\langle . \rangle$ is the time average. Readouts with greater SNR are less affected by noise than readouts with lower SNR.

\subsection*{Nonlinearity Metric}
The nonlinearity metric used in this study is defined in Equation \ref{eq:nonlinearity}. In Equation \ref{eq:nonlinearity}, $\chi_i$ is the frequency content of the signal at the $i^{th}$ frequency. The numerator is the sum over the set $\mathcal{N}$ of nonlinear frequencies (i.e., frequencies that are not found in the input signal), while the denominator is the sum over all frequencies.
\begin{equation} 
    \label{eq:nonlinearity}
    \nu=\frac{\sum_{i \in \mathcal{N}}\chi_i}{\sum_{i}\chi_i}
\end{equation}

\subsection*{Memory Metric}

First, we calculate the correlation function of the signal Y with respect to the four-tone input signal $I_4(t)$ delayed by $\tau$ time steps, i.e., $r(Y,I_4,\tau)$. The Pearson correlation between two signals Y and $I_4$ is defined as:
\begin{equation}
r(Y,I_4, \tau)=\frac{\sum_{i=\tau+1}^{N_T}(y_i-\bar{y})(I_{4,i-\tau}-\bar{I_4})}{\sqrt{\sum_{i=\tau+1}^{N_T}(y_i-\bar{y})^2\sum_{i=\tau+1}^{N_T}(I_{4,i-\tau}-\bar{I_4})^2}}
\end{equation}

where $N_T$ is the number of time steps in the signals, $y_i$ and $I_{4,i}$ are the $i^{th}$ elements in the signals Y and $I_4$, respectively, and $\bar{y}$ and $\bar{I_4}$ are the mean values of Y and $I_4$, respectively. The optimal time delay is calculated as:
\begin{equation}
    \tau_{opt} = \underset{\tau \le 25}{\text{argmax}}(|r(Y,I_4,\tau)|) 
\end{equation}
A cutoff of $\tau = 25$ time steps was used as the correlation function generally decayed over this interval, but then increased for $\tau > 25$ due to periodicity in the input signal. Once $\tau_{opt}$ was identified, the memory metric was calculated as:
\begin{equation}
    M = \tau_{opt} r(Y,I_4,\tau_{opt})
\end{equation}

\section*{Acknowledgments}
We acknowledge David M. Boston for his contribution on the Abaqus simulations of the metamaterial unit cells to support this study. We acknowledge the support of the Air Force Research Laboratory and the Air Force Office of Scientific Research, Grant \# FA9550–23–1–0519 and Grant \#26RXCOR021

\section*{Author Contributions}
S.H., S.K., P.B., and P.M. designed research; S.H. and S.K. performed research; S.H. and S.K. analyzed data; P.B., and P.M. supervision; and S.H., S.K., P.B., and P.M. wrote paper.

\section*{Competing Interests}
The authors declare no competing interest.

\section*{Corresponding Author}
\textsuperscript{1}To whom correspondence should be addressed. E-mail: pmusgrave@ufl.edu

\bibliographystyle{plain} 
\bibliography{pnas-citations}

@article{yasuda2021mechanical,
  title={Mechanical computing},
  author={Yasuda, Hiromi and Buskohl, Philip R and Gillman, Andrew and Murphey, Todd D and Stepney, Susan and Vaia, Richard A and Raney, Jordan R},
  journal={Nature},
  volume={598},
  number={7879},
  pages={39--48},
  year={2021},
  publisher={Nature Publishing Group UK London}
}

@article{raney2016stable,
  title={Stable propagation of mechanical signals in soft media using stored elastic energy},
  author={Raney, Jordan R and Nadkarni, Neel and Daraio, Chiara and Kochmann, Dennis M and Lewis, Jennifer A and Bertoldi, Katia},
  journal={Proceedings of the National Academy of Sciences},
  volume={113},
  number={35},
  pages={9722--9727},
  year={2016},
  publisher={National Academy of Sciences}
}

@article{coulais2016combinatorial,
  title={Combinatorial design of textured mechanical metamaterials},
  author={Coulais, Corentin and Teomy, Eial and De Reus, Koen and Shokef, Yair and Van Hecke, Martin},
  journal={Nature},
  volume={535},
  number={7613},
  pages={529--532},
  year={2016},
  publisher={Nature Publishing Group UK London}
}

@article{abdi2010principal,
  title={Principal component analysis},
  author={Abdi, Herv{\'e} and Williams, Lynne J},
  journal={Wiley interdisciplinary reviews: computational statistics},
  volume={2},
  number={4},
  pages={433--459},
  year={2010},
  publisher={Wiley Online Library}
}

@article{kiyabu2025optomechanical,
  title={Optomechanical reservoir computing},
  author={Kiyabu, Steven and Nelson, Daniel and Thomson, John and Schultz, Benjamin and Vincent, Timothy and Hertlein, Nathan and Gillman, Andrew and Criner, Amanda and Buskohl, Philip R},
  journal={Proceedings of the National Academy of Sciences},
  volume={122},
  number={29},
  pages={e2424991122},
  year={2025},
  publisher={National Academy of Sciences}
}

@article{inubushi2017reservoir,
  title={Reservoir computing beyond memory-nonlinearity trade-off},
  author={Inubushi, Masanobu and Yoshimura, Kazuyuki},
  journal={Scientific reports},
  volume={7},
  number={1},
  pages={10199},
  year={2017},
  publisher={Nature Publishing Group UK London}
}

@article{dambre2012information,
  title={Information processing capacity of dynamical systems},
  author={Dambre, Joni and Verstraeten, David and Schrauwen, Benjamin and Massar, Serge},
  journal={Scientific reports},
  volume={2},
  number={1},
  pages={514},
  year={2012},
  publisher={Nature Publishing Group UK London}
}

@inproceedings{maas2013rectifier,
  title={Rectifier nonlinearities improve neural network acoustic models},
  author={Maas, Andrew L and Hannun, Awni Y and Ng, Andrew Y and others},
  booktitle={Proc. icml},
  volume={30},
  number={1},
  pages={3},
  year={2013},
  organization={Atlanta, GA}
}

@article{jaeger2001short,
  title={Short term memory in echo state networks},
  author={Jaeger, Herbert},
  year={2001},
  publisher={GMD Forschungszentrum Informationstechnik}
}

@article{he2025role,
  title={The role of nonlinearity, dimensionality, memory, and input on a mechanical oscillator reservoir computer},
  author={He, Shan and Musgrave, Patrick F},
  journal={Neurocomputing},
  volume={652},
  pages={131158},
  year={2025},
  publisher={Elsevier}
}

@article{he2025physical,
  title={Physical reservoir computing on a soft bio-inspired swimmer},
  author={He, Shan and Musgrave, Patrick},
  journal={Neural Networks},
  volume={181},
  pages={106766},
  year={2025},
  publisher={Elsevier}
}

@article{bhovad2021physical,
  title={Physical reservoir computing with origami and its application to robotic crawling},
  author={Bhovad, Priyanka and Li, Suyi},
  journal={Scientific Reports},
  volume={11},
  number={1},
  pages={13002},
  year={2021},
  publisher={Nature Publishing Group UK London}
}

@article{bordiga2025nonlinear,
  title={Nonlinear mechanical metamaterial cloaks},
  author={Bordiga, Giovanni and Argaud, Jean-Gabriel and Watkins, Audrey A and Tournat, Vincent and Bertoldi, Katia},
  journal={Advanced Functional Materials},
  pages={e22895},
  year={2025},
  publisher={Wiley Online Library}
}

@article{tobalske2007biomechanics,
  title={Biomechanics of bird flight},
  author={Tobalske, Bret W},
  journal={Journal of Experimental Biology},
  volume={210},
  number={18},
  pages={3135--3146},
  year={2007},
  publisher={Company of Biologists}
}

@article{tanaka2019recent,
  title={Recent advances in physical reservoir computing: A review},
  author={Tanaka, Gouhei and Yamane, Toshiyuki and H{\'e}roux, Jean Benoit and Nakane, Ryosho and Kanazawa, Naoki and Takeda, Seiji and Numata, Hidetoshi and Nakano, Daiju and Hirose, Akira},
  journal={Neural Networks},
  volume={115},
  pages={100--123},
  year={2019},
  publisher={Elsevier}
}

@article{nakajima2013soft,
  title={A soft body as a reservoir: case studies in a dynamic model of octopus-inspired soft robotic arm},
  author={Nakajima, Kohei and Hauser, Helmut and Kang, Rongjie and Guglielmino, Emanuele and Caldwell, Darwin G and Pfeifer, Rolf},
  journal={Frontiers in computational neuroscience},
  volume={7},
  pages={91},
  year={2013},
  publisher={Frontiers Media SA}
}

@inproceedings{nelson2024spectral,
  title={Spectral Analysis of Mechanical Reservoir Computing With ReLU Spring Networks},
  author={Nelson, Daniel and Kiyabu, Steven and Vincent, Timothy and Gillman, Andrew and Criner, Amanda and Buskohl, Philip R},
  booktitle={Smart Materials, Adaptive Structures and Intelligent Systems},
  volume={88322},
  pages={V001T08A004},
  year={2024},
  organization={American Society of Mechanical Engineers}
}

@article{wang2024advancements,
  title={Advancements in soft robotics: a comprehensive review on actuation methods, materials, and applications},
  author={Wang, Yanmei and Wang, Yanen and Mushtaq, Ray Tahir and Wei, Qinghua},
  journal={Polymers},
  volume={16},
  number={8},
  pages={1087},
  year={2024},
  publisher={MDPI}
}

@article{eyvazian2026state,
  title={State-of-the-art soft robotic systems for unstructured and real-world environments: A systematic review},
  author={Eyvazian, Arameh and Song, Yooseob and Hovhannes, Chibukhchyan and Savari, Ardeshir and Singh, Narinderjit Singh Sawaran},
  journal={Engineering Science and Technology, an International Journal},
  volume={73},
  pages={102264},
  year={2026},
  publisher={Elsevier}
}

@article{mowla2025recent,
  title={Recent advancements in morphing applications: Architecture, artificial intelligence integration, challenges, and future trends-a comprehensive survey},
  author={Mowla, Md Najmul and Asadi, Davood and Durhasan, Tahir and Jafari, Javad Rashid and Amoozgar, Mohammadreza},
  journal={Aerospace Science and Technology},
  volume={161},
  pages={110102},
  year={2025},
  publisher={Elsevier}
}

@article{hoffmann2023bird,
  title={Bird-inspired robotics principles as a framework for developing smart aerospace materials},
  author={Hoffmann, Kenneth AW and Chen, Tony G and Cutkosky, Mark R and Lentink, David},
  journal={Journal of Composite Materials},
  volume={57},
  number={4},
  pages={679--710},
  year={2023},
  publisher={SAGE Publications Sage UK: London, England}
}

@article{aner2025decade,
  title={A Decade of Soft Robotic Manipulators: Advances in Design, Modeling, Control, and Emerging Challenges},
  author={Aner, Elsayed Atif and Shehata, Omar M and Awad, Mohammed Ibrahim and ElHady, Nancy E},
  journal={Journal of Bionic Engineering},
  pages={1--44},
  year={2025},
  publisher={Springer}
}

@article{wang2025metamaterial,
  title={Metamaterial beam with resonators of active feedback control to broaden and tune the bandgaps},
  author={Wang, Yuhang and Wang, Lifeng and Gao, Yuqiang},
  journal={Acta Mechanica},
  volume={236},
  number={3},
  pages={2331--2343},
  year={2025},
  publisher={Springer}
}

@article{poon2019phase,
  title={Phase-Changing Metamaterial Capable of Variable Stiffness and Shape Morphing},
  author={Poon, Ryan and Hopkins, Jonathan B},
  journal={Advanced Engineering Materials},
  volume={21},
  number={12},
  pages={1900802},
  year={2019},
  publisher={Wiley Online Library}
}

@article{haghpanah2016programmable,
  title={Programmable elastic metamaterials},
  author={Haghpanah, Babak and Ebrahimi, Hamid and Mousanezhad, Davood and Hopkins, Jonathan and Vaziri, Ashkan},
  journal={Advanced Engineering Materials},
  volume={18},
  number={4},
  pages={643--649},
  year={2016},
  publisher={Wiley Online Library}
}

@article{rus2015design,
  title={Design, fabrication and control of soft robots},
  author={Rus, Daniela and Tolley, Michael},
  journal={Nature},
  year={2015}
}

@article{ma2019review,
  title={A review of smart materials for morphing structures},
  author={Ma, Y.},
  journal={Composite Structures},
  year={2019}
}

@article{kim2013soft,
  title={Soft robotics: a bioinspired evolution in robotics},
  author={Kim, Sangbae},
  journal={Trends in Biotechnology},
  year={2013}
}

@article{maass2002real,
  title={Real-time computing without stable states: A new framework for neural computation based on perturbations},
  author={Maass, Wolfgang and Natschl{\"a}ger, Thomas and Markram, Henry},
  journal={Neural computation},
  volume={14},
  number={11},
  pages={2531--2560},
  year={2002},
  publisher={MIT Press}
}

@article{jaeger2004harnessing,
  title={Harnessing nonlinearity: Predicting chaotic systems and saving energy in wireless communication},
  author={Jaeger, Herbert and Haas, Harald},
  journal={science},
  volume={304},
  number={5667},
  pages={78--80},
  year={2004},
  publisher={American Association for the Advancement of Science}
}

@article{coulombe2017computing,
  title={Computing with networks of nonlinear mechanical oscillators},
  author={Coulombe, Jean C and York, Mark CA and Sylvestre, Julien},
  journal={PloS one},
  volume={12},
  number={6},
  pages={e0178663},
  year={2017},
  publisher={Public Library of Science San Francisco, CA USA}
}

@article{liang2024physical,
  title={Physical reservoir computing with emerging electronics},
  author={Liang, Xiangpeng and Tang, Jianshi and Zhong, Yanan and Gao, Bin and Qian, He and Wu, Huaqiang},
  journal={Nature Electronics},
  volume={7},
  number={3},
  pages={193--206},
  year={2024},
  publisher={Nature Publishing Group UK London}
}

@inproceedings{nowshin2020recent,
  title={Recent advances in reservoir computing with a focus on electronic reservoirs},
  author={Nowshin, Fabiha and Zhang, Yuhao and Liu, Lingjia and Yi, Yang},
  booktitle={2020 11th International Green and Sustainable Computing Workshops (IGSC)},
  pages={1--8},
  year={2020},
  organization={IEEE}
}

@article{rafayelyan2020large,
  title={Large-scale optical reservoir computing for spatiotemporal chaotic systems prediction},
  author={Rafayelyan, Mushegh and Dong, Jonathan and Tan, Yongqi and Krzakala, Florent and Gigan, Sylvain},
  journal={Physical Review X},
  volume={10},
  number={4},
  pages={041037},
  year={2020},
  publisher={APS}
}

@article{bu2022efficient,
  title={Efficient optical reservoir computing for parallel data processing},
  author={Bu, Ting and Zhang, He and Kumar, Santosh and Jin, Mingwei and Kumar, Prajnesh and Huang, Yuping},
  journal={Optics Letters},
  volume={47},
  number={15},
  pages={3784--3787},
  year={2022},
  publisher={Optica Publishing Group}
}

@inproceedings{zhang2022harnessing,
  title={Harnessing physical reservoir computing in nonlinear mechanical metastructures},
  author={Zhang, Yuning and Wang, Kon-Well},
  booktitle={AIAA Scitech 2022 Forum},
  pages={1464},
  year={2022}
}

@article{ameduri2023morphing,
  title={Morphing wings review: Aims, challenges, and current open issues of a technology},
  author={Ameduri, Salvatore and Concilio, A},
  journal={Proceedings of the Institution of Mechanical Engineers, Part C: Journal of Mechanical Engineering Science},
  volume={237},
  number={18},
  pages={4112--4130},
  year={2023},
  publisher={SAGE Publications Sage UK: London, England}
}

@article{vincent2025fluid,
  title={Fluid-based reservoir computing for distributed signal processing},
  author={Vincent, Timothy and Pankonien, Alex},
  journal={Journal of Intelligent Material Systems and Structures},
  volume={36},
  number={18-19},
  pages={1297--1300},
  year={2025},
  publisher={SAGE Publications Sage UK: London, England}
}

@inproceedings{vincent2026information,
  title={Information Processing Dynamics With Perturbed Cylinder-Flap Fluid-Structure Interactions},
  author={Vincent, Timothy and Boston, David M and Kiyabu, Steven and Pankonien, Alexander M and Buskohl, Philip},
  booktitle={AIAA SCITECH 2026 Forum},
  pages={0900},
  year={2026}
}

@article{kortman2025perspectives,
  title={Perspectives on intelligence in soft robotics},
  author={Kortman, Vera Gesina and Mazzolai, Barbara and Sakes, Aime{\'e} and Jovanova, Jovana},
  journal={Advanced Intelligent Systems},
  volume={7},
  number={1},
  pages={2400294},
  year={2025},
  publisher={Wiley Online Library}
}

@article{lee2022mechanical,
  title={Mechanical neural networks: Architected materials that learn behaviors},
  author={Lee, Ryan H and Mulder, Erwin AB and Hopkins, Jonathan B},
  journal={Science Robotics},
  volume={7},
  number={71},
  pages={eabq7278},
  year={2022},
  publisher={American Association for the Advancement of Science}
}

@article{nakajima2020physical,
  title={Physical reservoir computing—an introductory perspective},
  author={Nakajima, Kohei},
  journal={Japanese Journal of Applied Physics},
  volume={59},
  number={6},
  pages={060501},
  year={2020},
  publisher={IOP Publishing}
}

@article{goto2021twin,
  title={Twin vortex computer in fluid flow},
  author={Goto, Ken and Nakajima, Kohei and Notsu, Hirofumi},
  journal={New Journal of Physics},
  volume={23},
  number={6},
  pages={063051},
  year={2021},
  publisher={IOP Publishing}
}

@article{zhang2023embodying,
  title={Embodying Multifunctional Mechano-Intelligence in and Through Phononic Metastructures Harnessing Physical Reservoir Computing},
  author={Zhang, Yuning and Deshmukh, Aditya and Wang, Kon-Well},
  journal={Advanced Science},
  volume={10},
  number={34},
  pages={2305074},
  year={2023},
  publisher={Wiley Online Library}
}

\clearpage
\onecolumn

\begin{center}
{\Large Supporting Information: Embodying Intelligence into Mechanical Metamaterials via Reservoir Computing}
\end{center}

\vspace{0.5cm}

\setcounter{figure}{0}
\setcounter{table}{0}
\setcounter{equation}{0}
\setcounter{section}{0}

\renewcommand{\thefigure}{S\arabic{figure}}
\renewcommand{\thetable}{S\arabic{table}}
\renewcommand{\theequation}{S\arabic{equation}}
\renewcommand{\thesection}{S\arabic{section}}

\section*{Finite Element Model of Metamaterial Unit Cells}
\begin{figure*}[htbp]
    \centering
    \includegraphics[width=\textwidth]{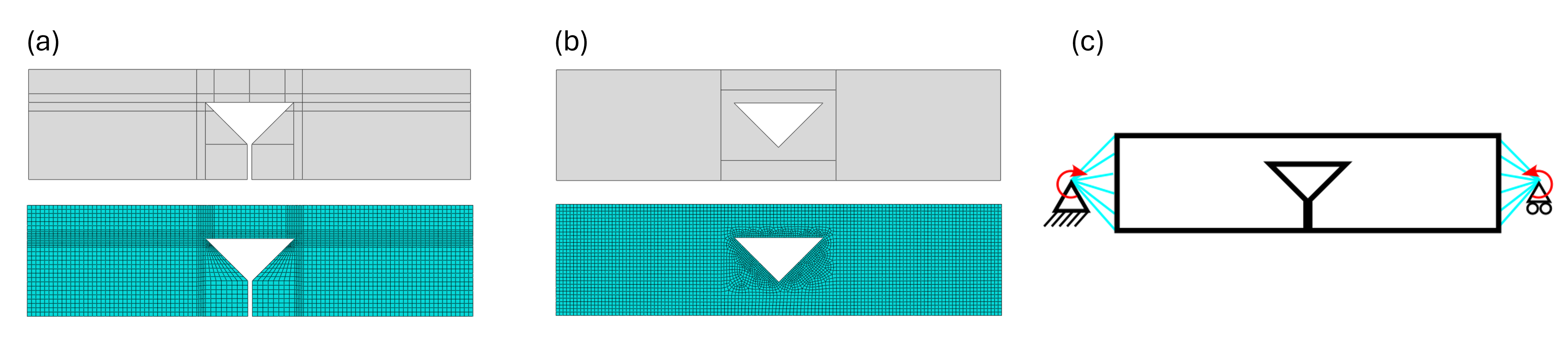}
    \caption{Geometric partitioning and mesh for (a) the nonlinear and (b) the linearized beam.
    (c) Schematic illustration of the boundary conditions for the static bending test.
    Light blue lines indicate kinematic coupling of the edges to dummy nodes.}
    \label{fig:FAE_beams}
\end{figure*}

The finite element analysis (FEA) of the ReLU beam is carried out using the Abaqus finite element software (Dassault Systems SIMULIA).
For simplicity, the model will be treated with the plane strain assumption, and only a 2-D projection of the nonlinear and linear geometries are considered.
The geometries are partitioned in both cases to allow for mesh refinement close to the corners of the triangular cutout in the beam (Figure \ref{fig:FAE_beams}-a-top).
In the nonlinear case (Figure \ref{fig:FAE_beams}-a), a structured quadrilateral mesh was suitable throughout, with an approximate element size of 0.1 mm in the fine regions and 0.25 mm elsewhere, resulting in a mesh of 3420 elements.
In the linear case (Figure \ref{fig:FAE_beams}-b), a transition region 0.75 mm away from the corners is partitioned.
Element nodes are seeded globally in the structured region with an approximate element size of 0.2 mm, while the transition region has mesh seeds along the cutout edges with approximate size of 0.1 mm.
The transition region is then meshed using Abaqus’s advancing front algorithm. This results in a mesh of 4577 elements.
All elements in both meshes are plane strain, reduced integration quadrilaterals (CPE4R in Abaqus).

A linear elastic material model is used, with an elastic modulus of 9.8 MPa and a Poisson’s ratio of 0.3.
Geometric nonlinearity is accounted for by Abaqus’s large-displacement formulation. A static bending test is carried out by imposing rotational boundary conditions at the midpoint of the beams’ vertical edges.
Because the elements are 2-D shells, they do not have an active rotational degree of freedom.
To reproduce this effect, a dummy node is created at the midpoints and coupled kinematically to the edge nodes, allowing the rotation of the dummy node to affect the translation of the edges (Figure \ref{fig:FAE_beams}-c).
To prevent rigid body motion, the left dummy node has all translation fixed, while the right node is restricted in the transverse direction, but free in the axial direction.
A $\pm10^\circ$ rotation is then imposed on the dummy nodes ($\pm15^\circ$ for the nonlinear tensile case) in 100 increments for each direction.
For each rotation increment, a reaction moment is extracted from each of the dummy nodes, subtracted from one another, and compared to the rotation to form the curves in Figure 2 in Main Body.

\section*{Experimental Characterization of ReLU Unit Cells}
\begin{figure*}[ht!]
    \centering
    \includegraphics[width=\textwidth]{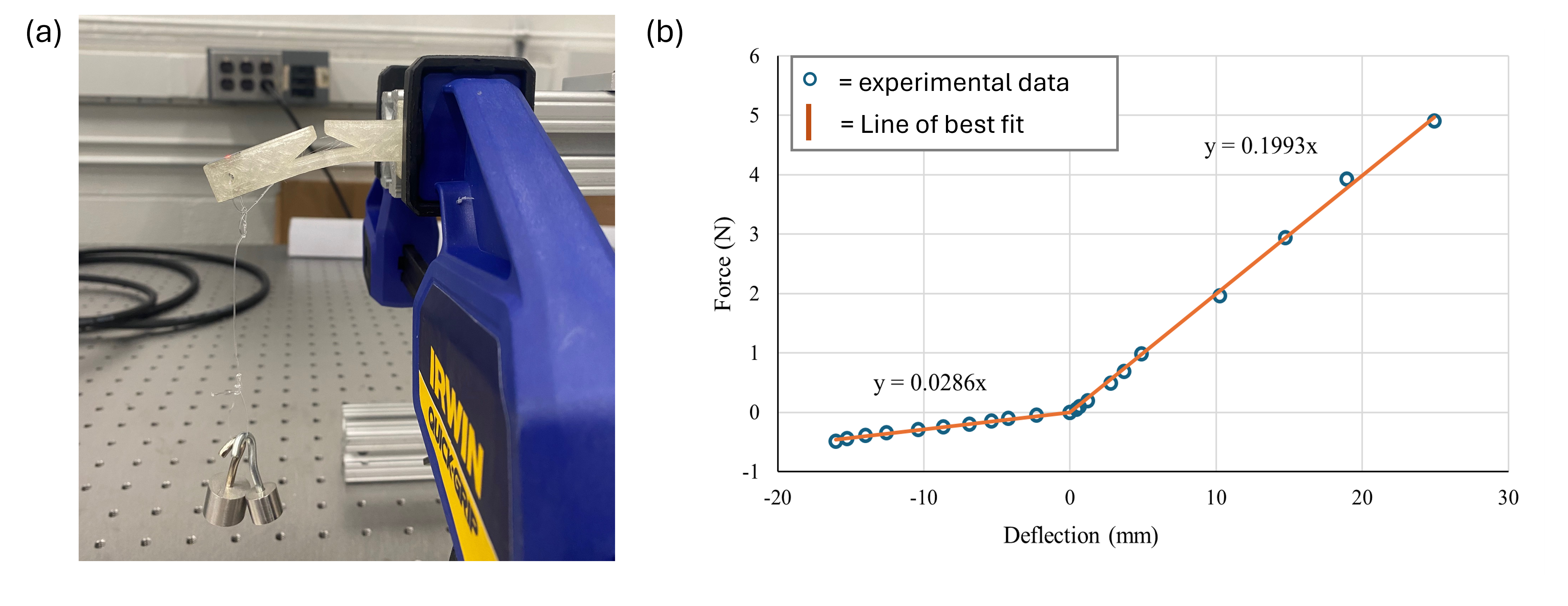}
    \caption{Static analysis of a single ReLU unit cell beam.
    (a) Experimental setup for static test: measuring displacement by adding weights using Laser Doppler Vibrometer (LDV) (b) Experimental data and its least-square fit of the bending stiffness.}
    \label{fig:ReLU_beam_analysis}
\end{figure*}

To characterize the bilinear bending stiffness of the nonlinear metamaterial, a static test was conducted on a single unit-cell beam.
A unit cell with identical geometry to those in the metamaterial was 3D-printed and mounted in a clamped-free boundary condition by fixing one end to an 80-20 structure.
A fishing line was attached to the free end, and incremental weights were applied to generate static input force, shown in Figure \ref{fig:ReLU_beam_analysis}-a.
A laser Doppler Vibrometer was used to measure the beam deflection under each applied load.
The data during the transient response immediately after loading were excluded, and only the measurements during equilibrium state after vibration had fully damped were extracted and averaged to obtain the deflection values.
A total of 21 data points were collected including 10 loading conditions in each bending direction and one zero-load case, shown in Figure \ref{fig:ReLU_beam_analysis}-b.
A linear least square fitting was then applied separately to each bending direction to determine the corresponding stiffness.
The results show distinct slopes for the positive (stiff) and negative (compliant) bending direction, confirming the bilinear stiffness behavior of the nonlinear metamaterial.

\section*{PCA of Nonlinear vs. Linear Metamaterial}
\begin{figure*}[ht!]
    \centering
    \includegraphics[width= 8cm]{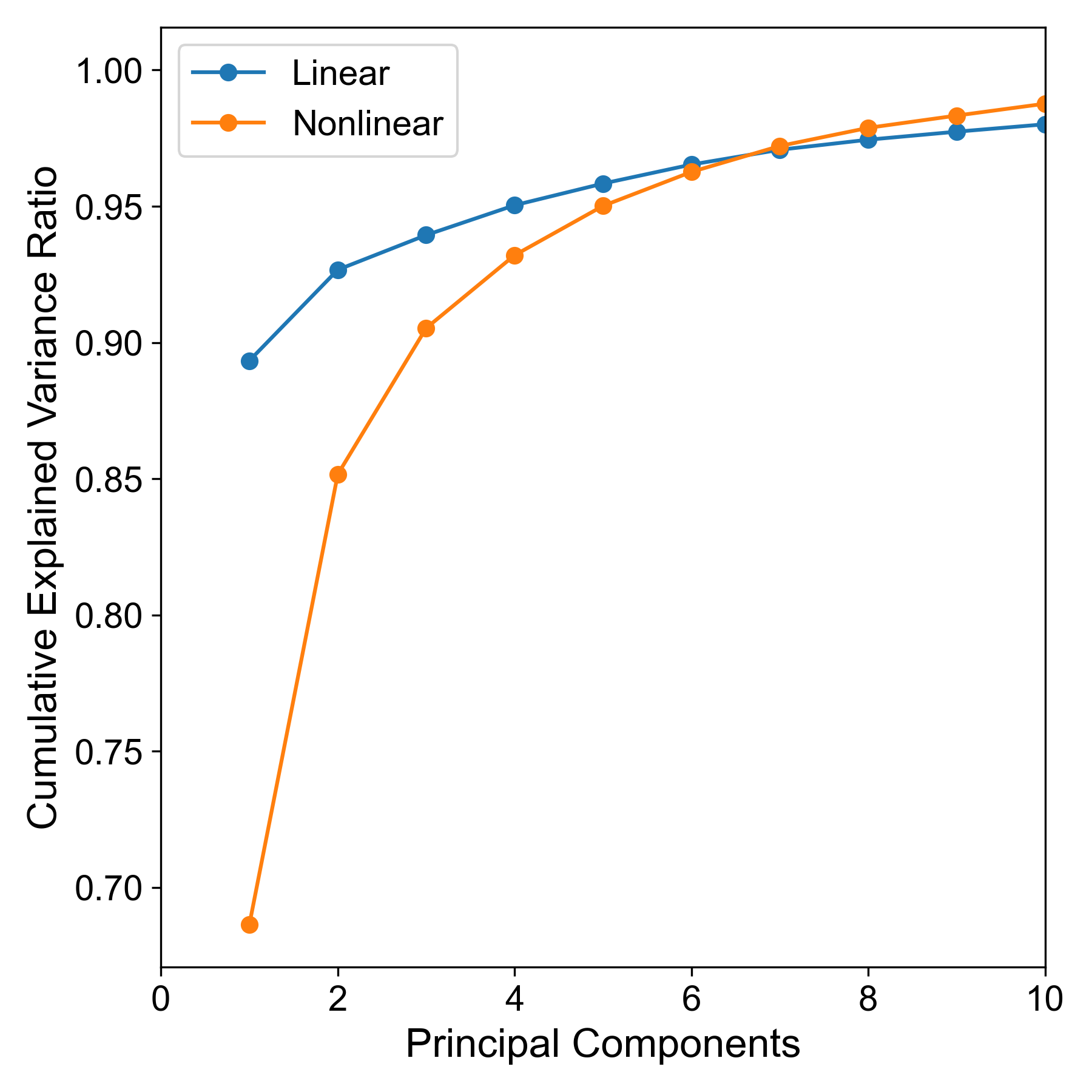}
    \caption{PCA analysis comparison between nonlinear metamaterial and linear metamaterial counterpart.
    Linear metamaterial exhibits high variance with fewer principal components, indicating most of readout information is explained using only fewer dimensions.
    Nonlinear metamaterial has low variance under fewer PCs but increase variance with larger number of PCs, indicating the readout information is well separated in a higher dimensions.}
    \label{fig:Explained_Variance_Lin_Nonlin}
\end{figure*}

To demonstrate that the nonlinear metamaterial exhibits richer and more separated dynamic information than its linearized counterpart, we performed principal component analysis (PCA) on both systems separately and compared their variance distributions.
Figure \ref{fig:Explained_Variance_Lin_Nonlin} shows the cumulative explained variance as a function of the number of principal components for the linear and nonlinear metamaterials.
In the linear system, the first principal component accounts for nearly 90\% of the system's dynamic variance, and the first ten components capture approximately 98\% of the system's dynamic variance.
In contrast, the nonlinear system exhibits a more distributed representation, with the first principal component explaining only 69\% of its variance.
The cumulative variance increases rapidly with additional components are included, reaching approximately 99\% of the total dynamic variance explained by the first ten principal components.
This difference highlights a key distinction in how the information is distributed in these two systems.
In the linear metamaterial, the dominance of the first few principal components indicates that most of the dynamic information is concentrated in a low-dimensional subspace, suggesting a limited information separation.
Conversely, in the nonlinear metamaterial, the variance is more distributed across multiple components, as evidenced by the larger slope of accumulation of explained variance.
This large slope of distributed variance indicated that the nonlinear system's dynamics span in a higher-dimensional space, suggesting well separation of information in the readout signals.

\section*{Sensor Selection for Independent Task}
\begin{figure*}[ht!]
    \centering
    \includegraphics[width=\textwidth]{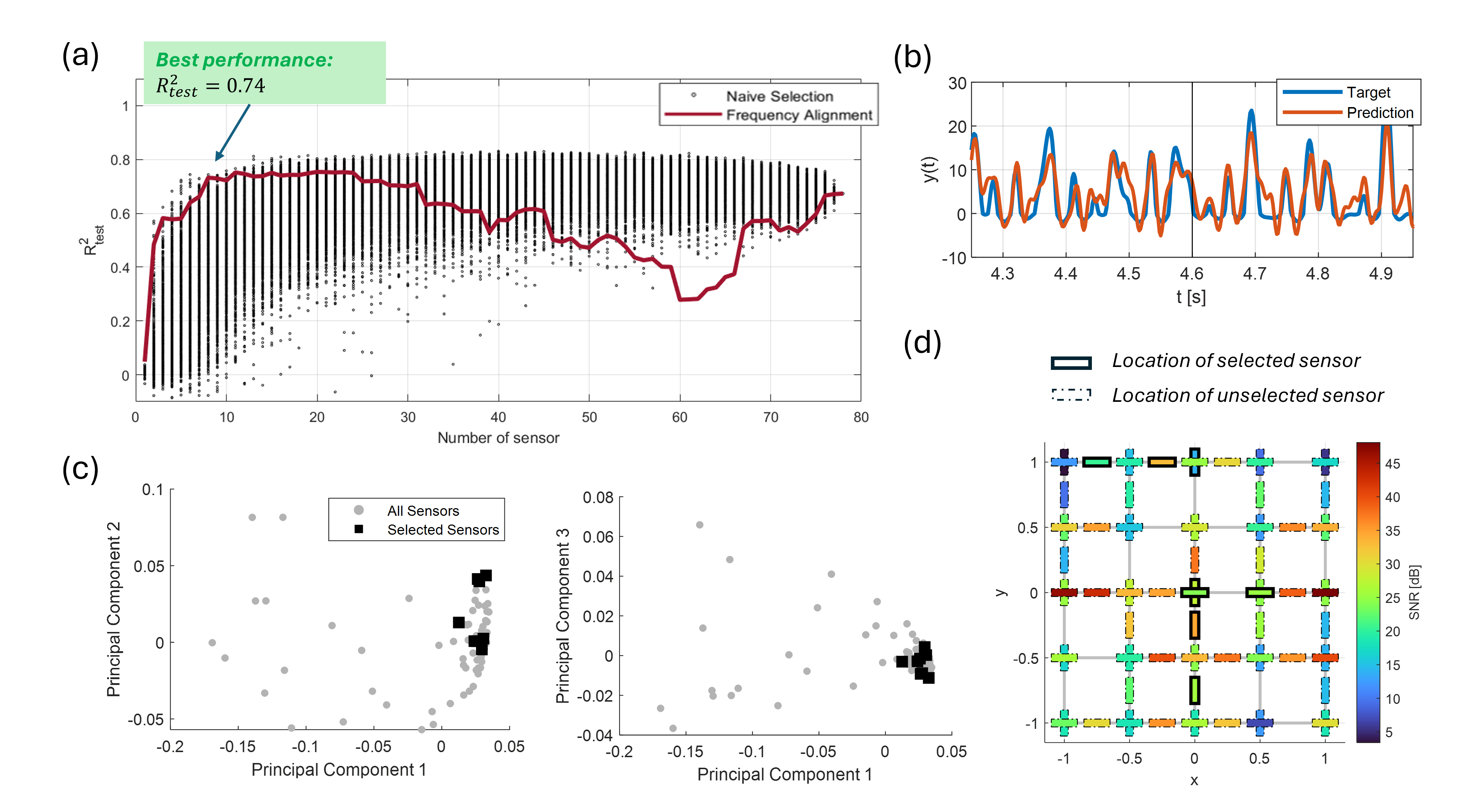}
    \caption{Prediction performance of independent ReLU-10 task as a function of readout sensors under a four-tone input case.
    (a) $R^2$ values for the ReLU-10 prediction task as sensors are incrementally added using a frequency alignment method (red solid line).
    For comparison, performance obtained from naïve sensor selection is shown using 1000 different random sensor combinations (black dots).
    Only eight sensors are used to achieve the highest $R^2$ value of 0.74 compared to the complete set of 78 sensors with $R^2$ value of 0.70.
    (b) Output signals of the embodied task using the 8 selected readout sensors, for the independent ReLU-10 task.
    (c) The selected 8 readout sensors highlighted in the PC-space, for the independent ReLU-10 task.
    (d) Physical locations of the readout sensors with their signal-to-noise ratio in dB.
    The eight selected sensor locations for the independent ReLU-10 tasks are indicated using thick solid marker line.
    Horizontal rectangular markers denote $\epsilon_{xx}$ and vertical markers denote $\epsilon_{yy}$}
    \label{fig:sensor selection_ReLu10}
\end{figure*}

Figure \ref{fig:sensor selection_ReLu10}-a shows the PRC performance for the independent ReLU-10 task using both the greedy sensor selection algorithm and naive sensor selection.
The solid curve represents the test performance achieved with the greedy selection algorithm.
For comparison, naive sensors selection using the same number of sensors is evaluated over 1,000 randomly sampled combinations from the pool of 78 sensors, shown as black dots.
The frequency-alignment method using the greedy algorithm converges to peak performance using only eight sensors, achieving $R^2=0.74$ and consistently outperforming the naive sensor approach.
In contrast, the prediction performance degrades as additional sensors are included.
The corresponding time-series prediction using the selected eight sensors is shown in Figure \ref{fig:sensor selection_ReLu10}-b.
These selected eight sensors highlghted in the principal component (PC) space is shown in Figure \ref{fig:sensor selection_ReLu10}-c, and their physical locations are shown in Figure \ref{fig:sensor selection_ReLu10}-d.

\end{document}